\def\kms{\hbox{km$\,$s$^{-1}$}}
\def\cmt{\hbox{cm$^{-3}$}}
\def\Apix{\hbox{\AA$\,$pix$^{-1}$}}
\def\two{\,{\sc ii}}

\def\farcs{\hbox{$.\!\!''$}}
\def\fsec{\hbox{$.\!\!^{\rm s}$}}

\documentclass{emulateapj}

\usepackage{amsmath}
\usepackage{amssymb}
\usepackage{mathrsfs}
\usepackage{color}
\usepackage{graphicx}

\shorttitle{Paper II: Nebular Properties of the M82 Starburst}
\shortauthors{M.\ S.\ Westmoquette et al.}

\begin{document}
\defcitealias{westm07c}{W07c}
\defcitealias{westm09a}{Paper I}
\defcitealias{westm09b}{Paper II} 


\title{The Optical Structure of the Starburst Galaxy M82. II. Nebular Properties of the Disk and Inner-Wind\altaffilmark{1}}
\author{M.\ S.\ Westmoquette\altaffilmark{2}\email{msw@star.ucl.ac.uk}}
\author{J.\ S.\ Gallagher\altaffilmark{3}}
\author{L.\ J.\ Smith\altaffilmark{2,4}}
\author{G.\ Trancho\altaffilmark{5}}
\author{N.\ Bastian\altaffilmark{6}}
\author{I.\ S.\ Konstantopoulos\altaffilmark{2}}

\altaffiltext{1}{Based on observations with the Gemini and WIYN telescopes}
\altaffiltext{2}{Department of Physics and Astronomy, University College London, Gower Street, London, WC1E 6BT, UK}
\altaffiltext{3}{Department of Astronomy, University of Wisconsin-Madison, 5534 Sterling, 475 North Charter St., Madison WI 53706, USA}
\altaffiltext{4}{Space Telescope Science Institute and European Space Agency, 3700 San Martin Drive, Baltimore, MD 21218, USA}
\altaffiltext{5}{Gemini Observatory, Casilla 603, Colina el Pino S/N, La Serena, Chile}
\altaffiltext{6}{Institute of Astronomy, University of Cambridge, Madingley Road, Cambridge, CB3 0HA, UK}
%

\begin{abstract}
In this second paper of the series, we present the results from optical Gemini-North GMOS-IFU and WIYN DensePak IFU spectroscopic observations of the starburst and inner wind zones of M82, with a focus on the state of the $T\sim10^4$~K ionized interstellar medium. Our electron density maps show peaks of a few 1000~\cmt\ (implying very high thermal pressures), local small spatial-scale variations, and a fall-off in the minor axis direction. We discuss the implications of these results with regards to the conditions/locations that may favour the escape of individual cluster winds that ultimately power the large-scale superwind. 

Our findings, when combined with the body of literature built up over the last decade on the state of the ISM in M82, imply that the starburst environment is highly fragmented into a range of clouds from small/dense clumps with low filling factors ($<$1~pc, $n_{\rm e}$$\gtrsim$$10^4$~\cmt) to larger filling factor, less dense gas. The most compact clouds seem to be found in the cores of the star cluster complexes, whereas the cloud sizes in the inter-complex region are larger. These dense clouds are bathed with an intense radiation field and embedded in an extensive high temperature ($T$\,$\gtrsim$\,$10^6$~K), X-ray-emitting ISM that is a product of the high star formation rates in the starburst zones of M82. The near-constant state of the ionization state of the $\sim$$10^4$~K gas throughout the M82 starburst zone can be explained as a consequence of the small cloud sizes, which allow the gas conditions to respond quickly to any changes. In Paper I we found that the observed emission lines are composed of multiple components, including a broad (FWHM $\sim$ 150--350~\kms) feature that we associate with emission from turbulent mixing layers (TMLs) on the surfaces of the gas clouds, resulting from the interaction of the fast wind outflows from the SSCs. The large number of compact clouds and wind sources provides an ideal environment for broad line emission, and explains the large observed broad/narrow line flux ratios. 

We have examined in more detail the discrete outflow channel identified within the inner wind in Paper~I. The channel appears as a coherent, expanding cylindrical structure of length $>$120~pc and and width 35--50~pc. The walls maintain an approximately constant (but subsonic) expansion velocity of $\sim$60~\kms, and are defined by peaks and troughs in the densities of the different line components. We hypothesise that as the hot wind fluid flows down the channel cavity, it interacts with the cooler, denser walls of the channel and with entrained material within the flow to produce broad-line emission, while the walls themselves emit primarily the narrow lines. We use the channel to examine further the relationship between the narrow and broad component emitting gas within the inner wind. Within the starburst energy injection zone, we find that turbulent motions (as traced by the broad component) appear to play an increasing role with height.

Finally, we have argued that a point-like knot identified in GMOS position 4, exhibiting blueshifted (by $\sim$140~\kms), broad ($\lesssim$350~\kms) H$\alpha$ emission and enhanced [S\two]/H$\alpha$ and [N\two]/H$\alpha$ ratios, is most likely an ejected luminous blue variable (LBV) type object.
\end{abstract}


\section{Introduction} \label{sect:intro}

Understanding feedback effects between massive stars, star clusters and the interstellar medium (ISM) is fundamental to the study of galaxy evolution. In this second paper of the series presenting high resolution optical integral field unit (IFU) observations of ionized gas in the starburst galaxy M82, we focus on the state of the nebular gas found in the disk and inner wind of this galaxy. This follows on from a companion paper \citep[][hereafter \citetalias{westm09a}]{westm09a} where we presented the data, described our observation and data reduction methods, and discussed the gaseous and stellar dynamics of the starburst core and inner disk of the galaxy.

M82 has long been known to be one of the nearest starburst galaxies with a large-scale optically-bright superwind \citep[see, e.g.,][]{lynds63}. This makes it of fundamental importance as an analogue to intensely star-forming galaxies at high-$z$. The current ($\sim$10~Myr) starburst activity is centred on the nucleus with a diameter of $\sim$500~pc ($\sim$30$''$). Within this region there are a number of prominent, high surface-brightness complexes\footnote{In this work we adopt the term starburst complexes \citep{bastian05} to describe regions of concentrated star/cluster formation in the M82 disk. We prefer this term to the one of starburst clumps used in Paper~I since this avoids confusion with gas clumps.} (labelled A--E), each known to contain many hundreds of young massive star clusters \citep{oconnell95, melo05}. It is the energy from these clusters that drives the outflow to kpc-scales \citep{shopbell98, stevens03a, strickland07}.

In \citetalias{westm09a} we found that the ionized gas within the starburst is dynamically very complex with much localised line splitting, and many overlapping expanding structures located at different radii. 
The high signal-to-noise of the data were sufficient for us to be able to accurately decompose the emission line profiles into multiple narrow components (FWHM $\sim$ 30--130~\kms) superimposed on a broad (FWHM $\sim$ 150--350~\kms) feature. Following the recent work by \citet[][the latter will hereafter be referred to as \citetalias{westm07c}]{westm07a, westm07b, westm07c}, we identified the broad component as originating from turbulent layers \citep{slavin93} on the surfaces of interstellar clouds set up by the impact of high energy photons and fast-flowing winds from the embedded young star clusters \citep[][see also \citealt{binette09}]{pittard05}. These layers may also be where mass is also stripped and entrained and/or loaded into the wind outflow \citep{suchkov94}. Results from \citetalias{westm09a} show that turbulent gas mixing into outflowing hot gas becomes increasingly important with height below the main disk of M82. Furthermore, careful inspection of the ionized gas properties directly below complex C revealed a discrete channel within the outflow. Further investigation of this interesting structure forms part of the focus of this paper.

High central densities, $n_{\rm e}$, (hence thermal pressures) and steep radial density gradients have long been known in M82 \citep{oconnell78, heckman90, mckeith95}, and formed one of the key pieces of evidence supporting the starburst/superwind explanation of the M82 ``explosion'' \citep[first reported by][]{lynds63}. On the most basic level, the collective effect of the supernovae and stellar winds in a starburst blows a bubble of hot gas that is driven outwards by a pressure much greater than the surrounding ambient medium. The overpressurisation of this bubble drives it to expand further and, if the energy input is sufficient, to blow-out of the galactic disk, allowing a coherent outflow of material to develop. The absolute value of the central density measured in M82 has, to some degree, been a function of the spatial resolution of the observations, since the nuclear regions of M82 exhibit a complex morphology of complexes, voids and dust lanes. From \textit{HST}/STIS spectroscopy, we measured peak densities in the range 1500--2000~\cmt\ \citepalias{westm07c}. Until now, densities have mostly been measured using one-dimensional slits (albeit at various positions and PAs), but this method is clearly limited in the amount of information available regarding the full two-dimensional variation of the densities.

Spatially resolved diagnostics of the gas excitation state, however, are easier to come by since, in the crudest form, two narrow-band images can simply be ratioed. Gas excitations have been measured from emission line ratios in the optical through to far-IR at a number of spatial resolutions (\citealt{satyapal95, achtermann95, lord96, shopbell98, forster01}; \citetalias{westm07c}; \citealt{beirao08}). The dominant finding of these studies is that there is very little variation in the line ratios within the starburst region, indicating a high degree of uniformity of conditions within the ionized gas. This is surprising considering the dramatic density variations and the existence of multiple, spatially defined, luminous starburst complexes.

Formed only with the last decade, the emerging picture of conditions within the M82 starburst has resulted from the comparison of multi-wavelength molecular and atomic gas observations with detailed PDR and photoionization codes. To summarise, the ISM is thought to be highly fragmented into a continuum of cloud sizes and densities (ranging from $<$1pc and $\lesssim$few $\times$ $10^4$~\cmt), with low area covering factors (\citealt{lord96, colbert99, forster01, forster03, r-r04}; \citetalias{westm07c}). A recent study by \citet{fuente08} indicated that $\sim$90\% of the mass of the molecular gas exists within small clouds with $A_{V}\gtrsim 5$~mag. These clouds are highly pressured, and partly disrupted and dissociated \citep{mao00, weis01} as a result of being in such a high energy environment. The clouds are embedded in a hot, diffuse surrounding medium of equal pressure \citep[recently detected and modelled by][]{strickland07, strickland09} maintained by the large number of SN shocks propagating through the starburst. In order to maintain the observed levels of ionization and excitation, \citeauthor{lord96} found the average separation of the clouds and ionizing sources had to be of order their size (1--7~pc). This agrees well with the average cluster separations catalogued by \citet{melo05} with \textit{HST} observations, and with high-resolution CO observations \citep{mao00, weis01}. Thus, the fragmented and well-mixed nature of the ISM within the starburst must allow it to rapidly equalise after environmental changes, explaining the aforementioned consistency of the line ratios \citepalias{westm07c}. 

\citet{shopbell98} and \citet{beirao08} found evidence for changes in the gas excitation out of the plane along outflow axis (minor axis). This has been interpreted as resulting from a combination of a decrease in gas density leading to an increase in the ionization parameter, dilution of radiation field, and the increasing importance of shocks in the halo. Studying the relationship between the gas conditions in the starburst and the outflow, and the transition between the two, is of essential importance for understanding how the wind develops and evolves.

In this paper we will focus on the derived nebular properties of the ionized gas, including the densities and excitations. By combining both detailed ($\sim$$0\farcs7$) observations of the starburst core and inner wind made with the Gemini GMOS-IFU, and wider-field observations of the inner $\sim$2~kpc of the disk made with the DensePak instrument, we will attempt to build up a coherent model of the ISM conditions (physical composition, densities, excitations) within the starburst, use the density distribution to discuss implications for the escape of individual cluster winds in different parts of the starburst, and examine the ionized gas flow properties within the inner wind and the starburst energy injection zone.


In this work, we have adopted a distance to M82 of 3.6~Mpc \citep{freedman94, mccommas09}, meaning $1'' = 17.5$~pc.

\section{Observations and data reduction} \label{sect:obs}

A full description of our observations and data reduction methods is given in \citetalias{westm09a}. In the following we will briefly recap some of the main points.

\subsection{Gemini observations} \label{sect:GMOS_obs}
\begin{figure*}
\centering
\includegraphics[width=\textwidth]{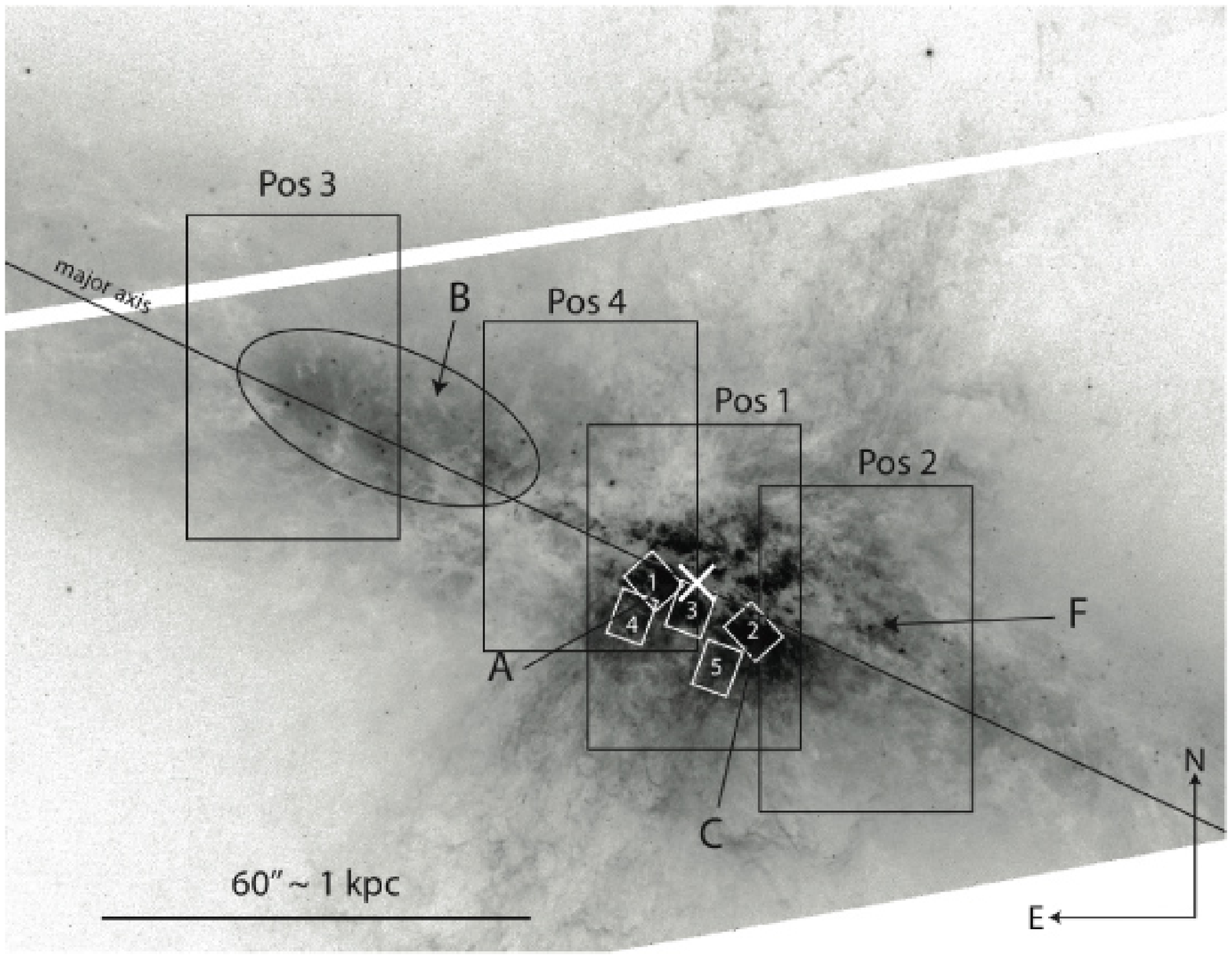} 
\caption{\textit{HST} F658N (H$\alpha$) image of M82 showing the positions of the GMOS (white) and DensePak (black) IFUs. Some of the bright starburst complexes/clusters are marked with letters \citep[nomenclature from][]{oconnell95}, the position of the 2.2~$\mu$m nucleus \citep{lester90} is shown with a white cross, and the galaxy major axis (PA = 65$^{\circ}$) is indicated by a solid line.}
\label{fig:finder}
\end{figure*}

\begin{figure*}
\centering
\includegraphics[width=0.95\textwidth]{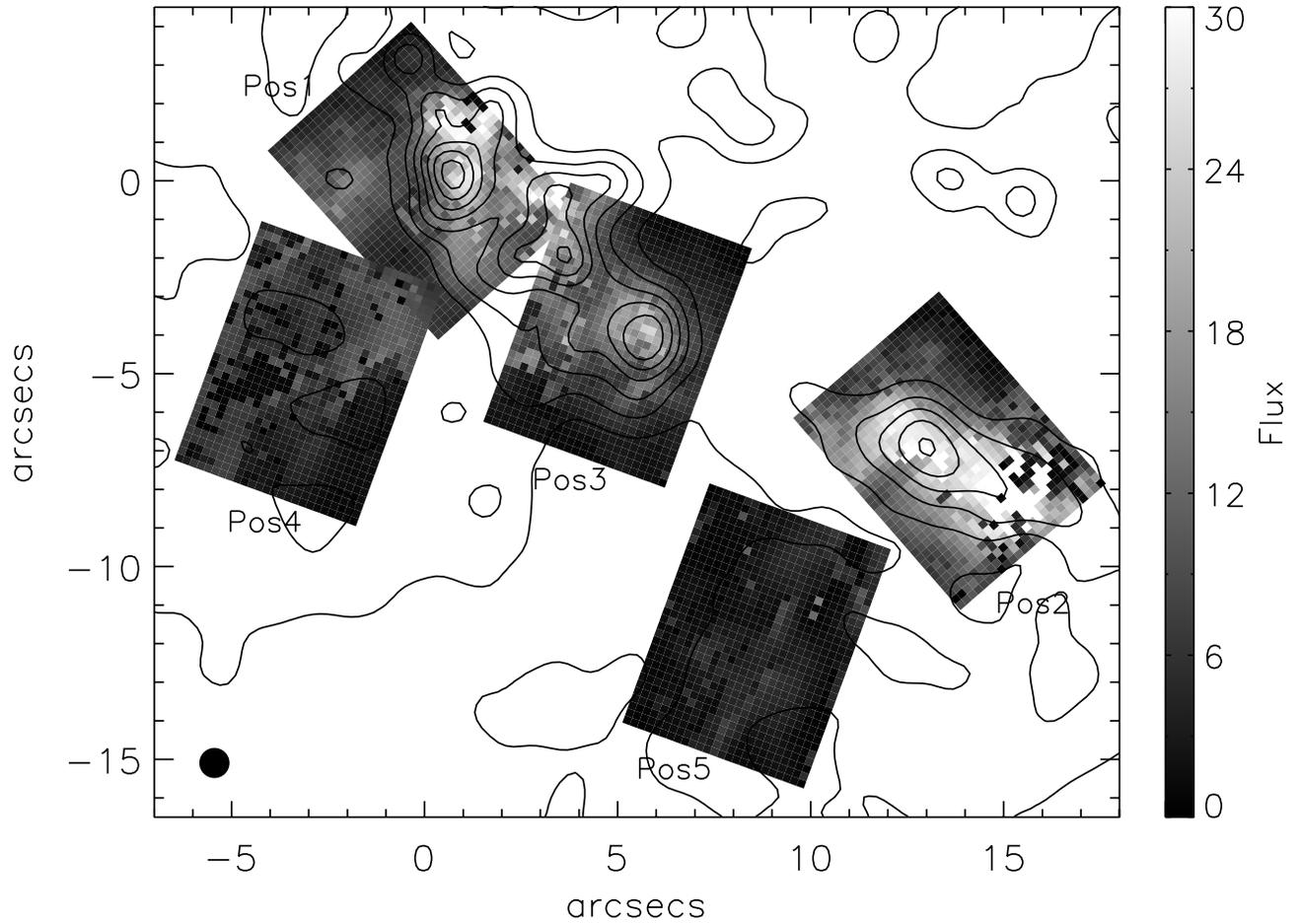} 
\caption{H$\alpha$ flux maps for C1. A scale bar is given for each map in arbitrary (but relative) flux units. Contours of the ACS F435W ($B$-band) are overplotted to illustrate the locations and extents of regions A and C. The maximum size of the seeing disk (0$\farcs8$) is indicated by the black circle in the lower-left and applies to all subsequent GMOS figures.}
\label{fig:Hac1_fl}
\end{figure*}

In February 2006 and February 2007 observations using the Gemini-North Multi-Object Spectrograph (GMOS) Integral Field Unit \citep[IFU;][]{allington02} were obtained covering five regions near the centre of M82 (programme IDs: GN-2006A-Q-38, and GN-2007A-Q-21, PI: L.J.\ Smith), with 0.3--0.8 arcsec seeing (5--14 pc at the distance of M82; see indicator on Fig.~\ref{fig:Hac1_fl}). The IFU in two-slit mode gives a field-of-view of $7\times 5$~arcsecs (which corresponds to approximately $50\times 35$~pc at the distance of M82) sampled by 1000 hexagonal contiguous fibres of $0\farcs2$ diameter. A separate block of sky fibres, offset by 1~arcmin from the object field, provides simultaneous sky observations. We took two dithered exposures at each position with integration times between 1200 and 2400~s each, and used the R831 grating to give a spectral coverage of 6100--6790~\AA\ with a dispersion of 0.34~\Apix. A spectral resolution of $R\simeq 5200$ was measured from single Gaussian fits to isolated arc lines. The locations of each field are shown on an \textit{HST}/ACS F658N image in Fig.~\ref{fig:finder}.

Basic data reduction was performed following the standard Gemini reduction pipeline (implemented in \textsc{iraf}\footnote{The Image Reduction and Analysis Facility ({\sc iraf}) is distributed by the National Optical Astronomy Observatories which is operated by the Association of Universities for Research in Astronomy, Inc. under cooperative agreement with the National Science Foundation.}). After reduction and combining the dithered exposures the five final data cubes each contained 825 wavelength and flux calibrated, sky subtracted spectra.

\subsection{WIYN observations} \label{sect:DP_obs}
DensePak \citep{barden98} was a small fibre-fed integral field array attached to the Nasmyth focus of the WIYN\footnote{The WIYN Observatory is a joint facility of the University of Wisconsin-Madison, Indiana University, Yale University, and the National Optical Astronomy Observatories.} 3.5-m telescope. It had 91 $3''$ fibres arranged in an array of $30\times 45$~arcsecs. Four additional fibres, offset by $\sim$$60''$ from the array centre, served as dedicated sky fibres.

On 14th April 2001, we used DensePak to observe the inner disk of M82. The 860~line~mm$^{-1}$ grating was used in two settings; four fields were observed in the wavelength range of 5820--6755~\AA\ (dispersion 0.46~\Apix; $R\simeq 6700$), three of these positions were also observed in the wavelength range of 7745--9700~\AA\ (dispersion 0.96~\Apix; $R\simeq 3650$). These settings allowed us to cover many of the brightest optical and near-IR (NIR) nebular emission and stellar absorption lines. A number of bias frames, flat-fields and arc calibration exposures were also taken together with the science frames. Basic reduction was achieved using standard {\sc iraf} packages. After reduction, the datafiles contained 86 wavelength calibrated and sky-subtracted spectra. Fig.~\ref{fig:finder} shows the position of the DensePak fields overlaid on the \textit{HST} F658N image.

\subsection{Decomposing the line profiles} \label{sect:line_profiles}

Following the methodology employed by \citet{westm07a,westm07b}, we fitted multiple Gaussian profile models to each emission line using an \textsc{idl} $\chi^{2}$-fitting package called \textsc{pan} \citep[Peak ANalysis;][]{dimeo}, to quantify the gas properties observed in each IFU field.

H$\alpha$ is detected in every fibre of both the GMOS and DensePak datasets. The high S/N and spectral resolution of our data allowed us to quantify the line profile shapes to a high degree of accuracy. In general, we find the emission lines to be composed of a bright, narrow component (hereafter referred to as C1; this was sometimes split into two narrow components, hereafter C1 and C3) overlaid on a fainter, broad component (hereafter C2). In some regions of our DensePak data, we can also identify an H$\alpha$ absorption component of stellar origin (hereafter C4). A full description of the procedure we employed to fit the lines and automatically determine the number of Gaussian models to fit is given in \citetalias{westm09a}, together with estimates of the fit uncertainties. It is prudent to recall that for the purposes of consistency in the fitting and analysis, where a triple-Gaussian fit was chosen we specified that the broadest component should be assigned to component 2 (C2), and after that, the brightest to be component 1 (C1), and the fainter narrow component to component 3 (C3).

Given the limitations of current IFU observing techniques and multi-component line fitting methods, a rigorous analysis of the uncertainties associated with the results presented in this paper is currently not feasible. Firstly, the flux calibration of IFU data is not just subject to the usual kinds of uncertainties found in spectroscopic data, but also to additional complications due to the complexity and design of the instrument. Furthermore, the combination and interpolation of the dithered IFU maps introduces additional uncertainties. These factors together mean that we cannot rigorously quantify the uncertainties in the calibration of the individual spectra. Secondly, although a reduced $\chi^2$ value for a multi-component line fit can be calculated, we currently do not fully understand how to quantify the uncertainties associated with each individual Gaussian component fit to the line profile. However, we can estimate errors from the observed variances in observed parameters in regions where conditions are relatively stable.

As discussed in \citetalias{westm09a}, we estimate that, for the fluxes, there is a 0.5--10 per cent error in C1 (with the range resulting mostly from the variation in S/N levels), 8--15 per cent on C2 and 10--80 per cent on the C3 flux, with the level depending on the relative flux in C3. The presence of a third fit component may slightly increase the errors in the C1 and C2 fluxes particularly where C3 is faint. When these estimates are carried through, we might expect similar percentage errors for the line ratio and electron density measurements. Due to the aforementioned difficulties, we do not attempt to include error maps or show error bars for each figure, but where specific measurements are quoted in the text we have made an effort to indicate approximate uncertainties. Finally, we have converted the fluxes of the H$\alpha$ absorption component (C4) to equivalent widths (EWs), and we estimate that the uncertainties on the EWs lie in the range of $\sim$0.5--1~\AA.

In \citetalias{westm09a}, we presented the line width and radial velocity results from the GMOS and DensePak datasets. In the sections following this, we will present and describe the remainder of our results, including the line flux and flux-derived properties of the emission line gas.

\section{Nebular properties of the nuclear regions and inner wind} \label{sect:GMOS}

\subsection{H$\alpha$ flux distribution} \label{sect:GMOS_Ha}

\begin{figure*}
\centering
\includegraphics[width=0.95\textwidth]{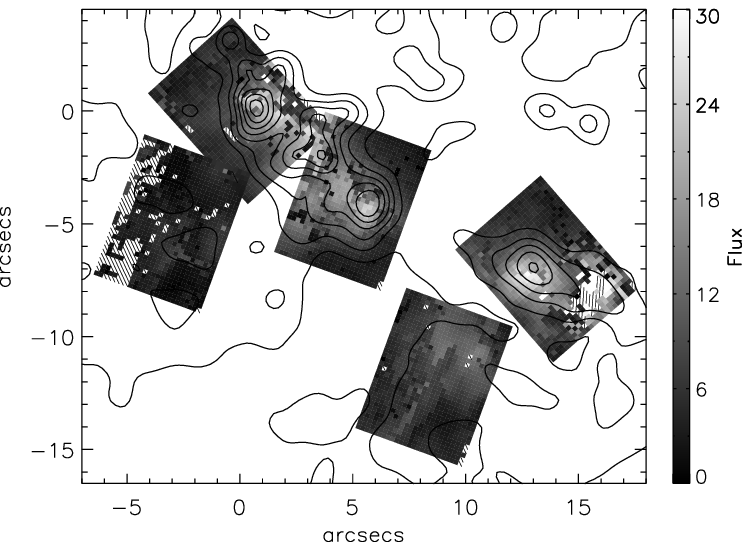} 
\caption{H$\alpha$ flux maps for C2. Labels as for Fig.~\ref{fig:Hac1_fl}.}
\label{fig:Hac2_fl}
\end{figure*}

\begin{figure*}
\centering
\includegraphics[width=0.95\textwidth]{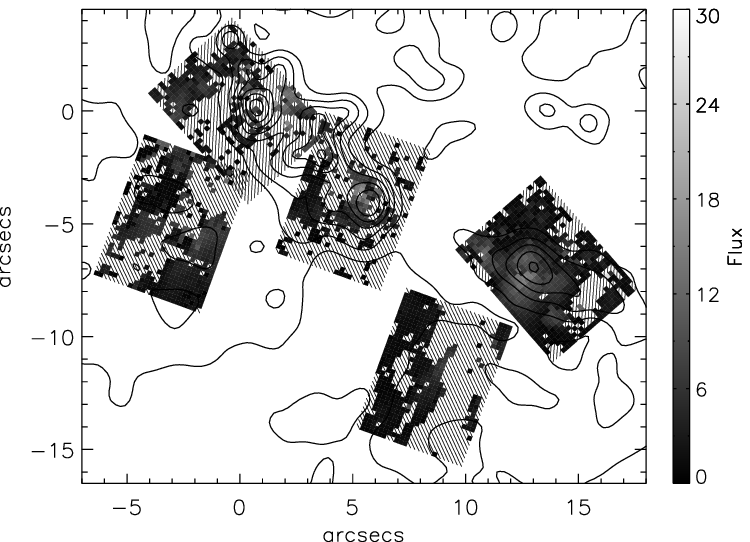} 
\caption{H$\alpha$ flux maps for C3. Labels as for Fig.~\ref{fig:Hac1_fl}.}
\label{fig:Hac3_fl}
\end{figure*}

Figs.~\ref{fig:Hac1_fl}, \ref{fig:Hac2_fl} and \ref{fig:Hac3_fl} show maps of the H$\alpha$ fluxes for the three line components, respectively. For reference, we have overplotted contours from the \textit{HST}/ACS F439W \citep[$\sim$$B$-band;][]{mutchler07} image in order to highlight the distribution of stellar emission and therefore the locations and extents of regions A and C. We have also indicated the maximum size of the seeing disk experienced during these observations ($0\farcs8$) with filled circle. We begin by noting that C1 and C2 are detected over almost all of the five positions, and C3 is identified in approximately one third of all spaxels. We also note that the typical EW of the H$\alpha$ emission line in position 1 is 50~\AA, meaning that we can safely ignore the effects of underlying stellar absorption.

To a first-order approximation the flux morphology in the three H$\alpha$ line components is fairly similar. However there are important differences. Firstly, we notice that the region around the cluster M82-A1 \citep[northern corner of position 1;][]{smith06} does not stand out in any of the three component maps (although it is clearly seen in the F439W contours). Secondly, the C2 and C3 fluxes appear to be brighter on the southern side of any prominent feature compared to C1. For instance in position 1 (region A) the flux of C1 peaks to the north and east of region A, whereas C2 peaks in the centre to south-west of region A (its flux, however, is strongly influenced by the varying contribution from C3). Furthermore, the peak in C1 flux coincident with the large H$\alpha$ knot in position 3 is located slightly to the north compared to the C2 peak. In position 2, this offset trend is again seen, with the peaks in C2 and C3 being offset slightly to the south compared to C1.
 
\begin{figure*}
\centering
\includegraphics[width=0.95\textwidth]{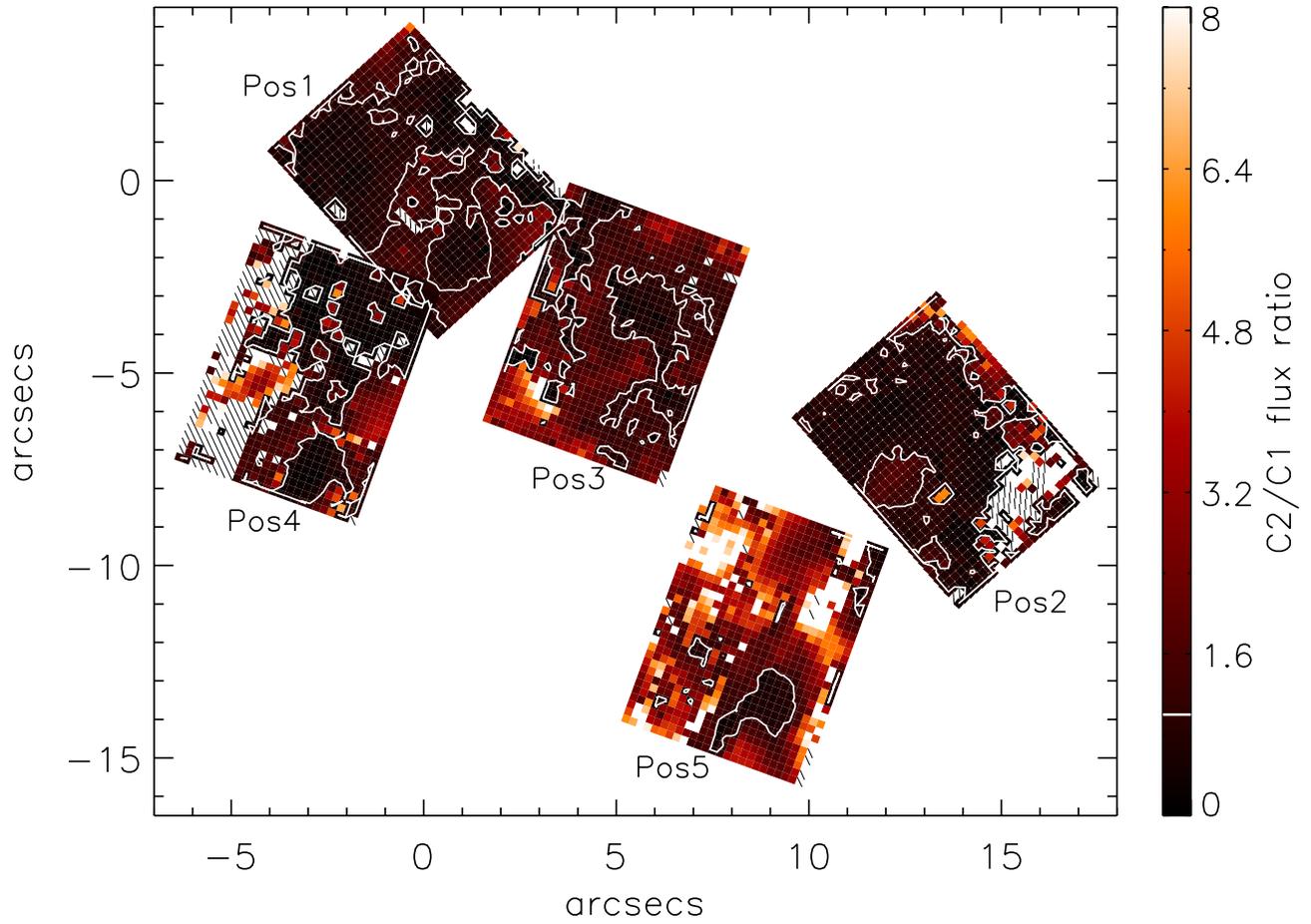} 
\caption{GMOS H$\alpha$ C2/C1 flux ratio maps. The single contour level represents a ratio of 1 (i.e.\ C2 = C1), as also indicated with a marker in the scale bar. \textit{(A color version of this figure is available in the online journal.)} }
\label{fig:Hac2c1_ratio}
\end{figure*}

In order to assess the importance of C2 in terms of its contribution to the total H$\alpha$ flux, we have mapped the the C2/C1 flux ratio for all five IFU positions in Fig.~\ref{fig:Hac2c1_ratio}. Surprisingly, of the spaxels that have both a C1 and C2 detection, $\sim$60\% exhibit C2/C1 flux ratios greater than unity (i.e.\ on average C2 contributes more to the total line flux than C1; see also Table~\ref{tbl:Ha_ratios}). Ratios $>$10 most likely result simply from bad fits. However, true ratios in the range 5--10 are seen in the wind positions 4 and 5 (fig.~4f of \citetalias{westm09a} shows an example line profile with a C2/C1 flux ratio in this range), and there are a significant number of spaxels with C2/C1 ratios of $\sim$2 in position 3 and parts of the south-west of position 1. The majority of spaxels with ratios $<$1 are found in positions 1 and 2 (regions A and C), whereas those exhibiting the lowest ratios (0.1--0.3) are found in position 4 \citepalias[coincident with the regions containing the broadest C2 emission, i.e.\ the position 4 knot -- see][]{westm09a}. In summary, although the C2/C1 flux ratio varies significantly and rapidly over the five IFU fields, the general trend is of increasing C2/C1 with radius. This is discussed further and in context with the rest of our findings in Section~\ref{sect:channel}. Finally, we note that finding such a large number of high ($>$1) C2/C1 flux ratios is in sharp contrast with what was found in NGC 1569 by \citet{westm07a, westm07b}. There, ratios fell in the range 0.1--0.3 with an average 0.12.

\subsection{Electron densities} \label{sect:GMOS_elecdens}
\begin{figure*}
\centering
\includegraphics[width=0.95\textwidth]{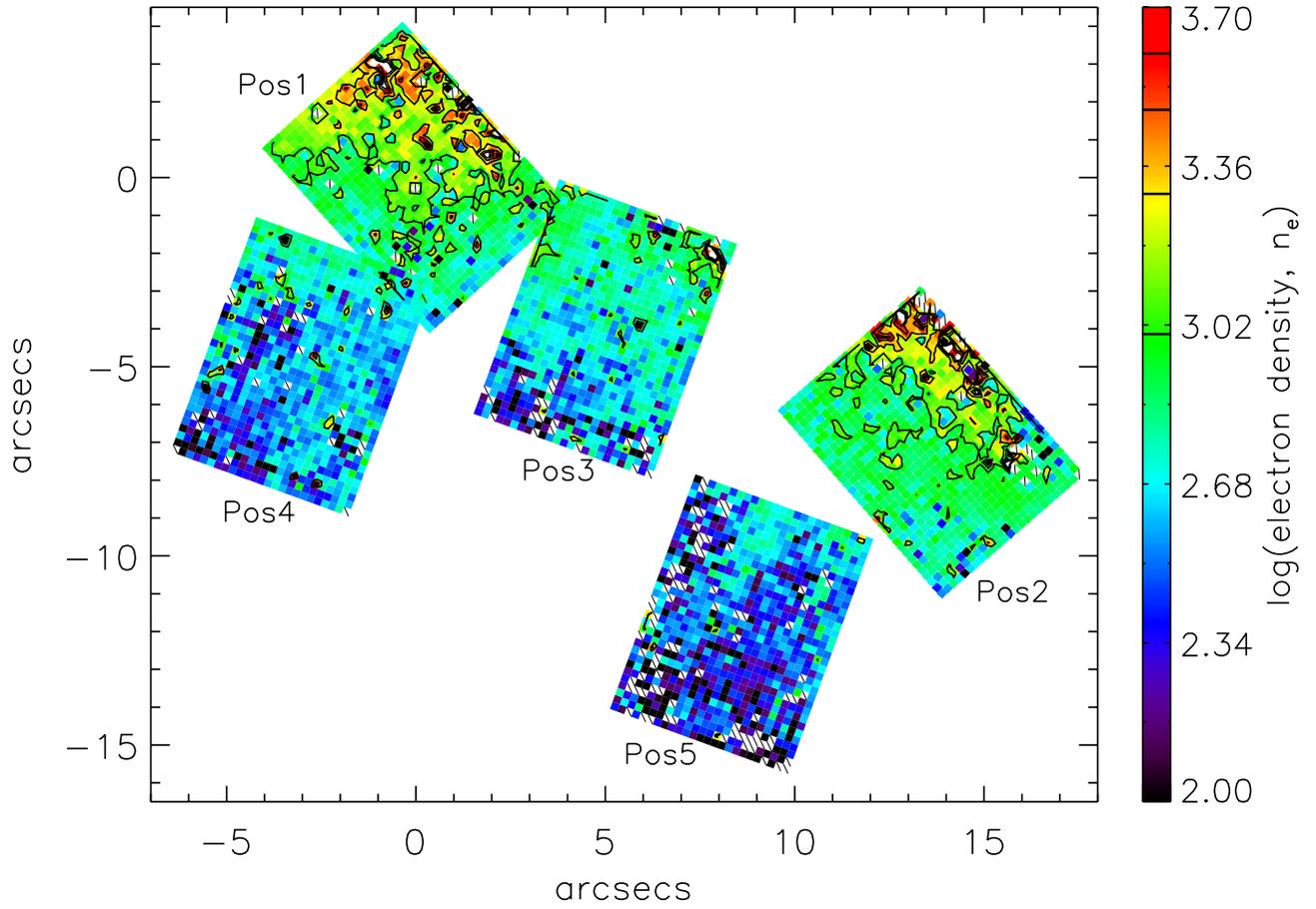} 
\caption{Maps of the [S\two]$\lambda\lambda$6717,6731 derived electron densities ($n_{\rm e}$, in units of \cmt), for C1. The lower limit to the colour range equates to 100~\cmt, the low-density limit of this indicator, and the contours represent 1000, 2000, 3000, and 4000~\cmt, as marked on the scale bars. \textit{(A color version of this figure is available in the online journal.)} }
\label{fig:elecdens_C1}
\end{figure*}

\begin{figure*}
\centering
\includegraphics[width=0.95\textwidth]{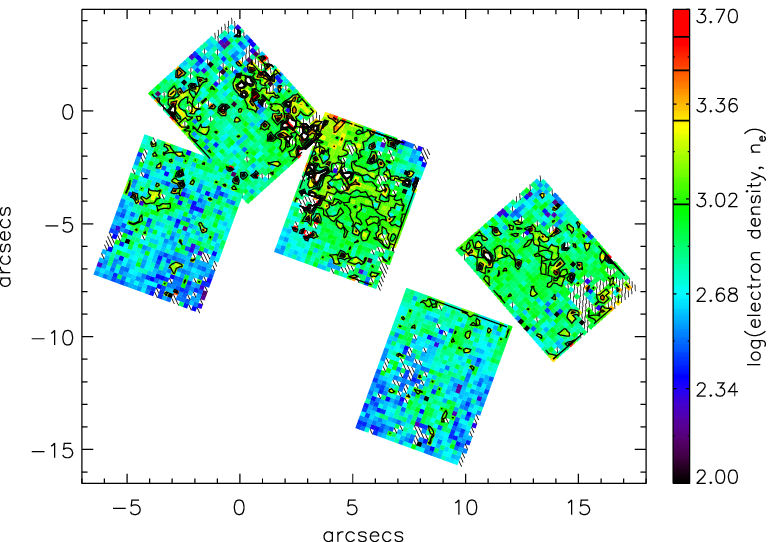} 
\caption{Maps of the electron densities for C2. Labels as for Fig.~\ref{fig:elecdens_C1}. \textit{(A color version of this figure is available in the online journal.)} }
\label{fig:elecdens_C2}
\end{figure*}

\begin{figure*}
\centering
\includegraphics[width=0.95\textwidth]{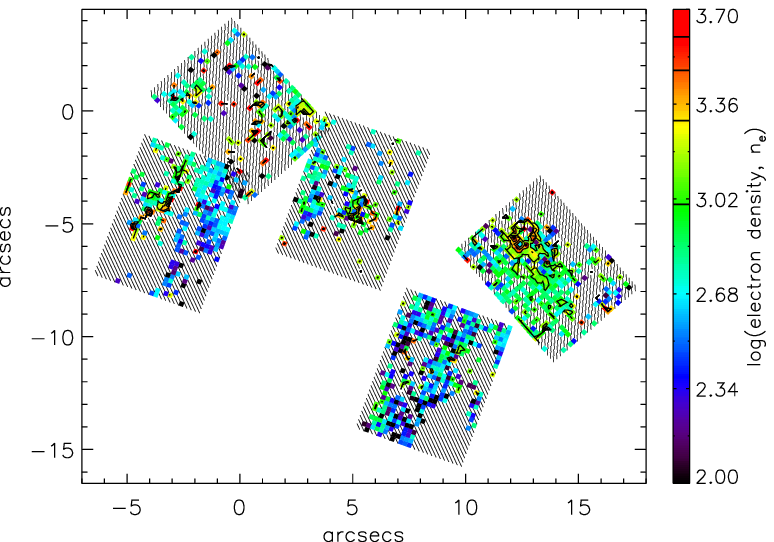} 
\caption{Maps of the electron densities for C3. Labels as for Fig.~\ref{fig:elecdens_C1}. \textit{(A color version of this figure is available in the online journal.)} }
\label{fig:elecdens_C3}
\end{figure*}

Figs.~\ref{fig:elecdens_C1}, \ref{fig:elecdens_C2} and \ref{fig:elecdens_C3} show maps of the spatial variation of electron density, $n_{\rm e}$, in each of the three line components, derived from flux ratios between Gaussian fits to the [S\two]$\lambda\lambda$6717,6731 lines (assuming an electron temperature, $T_{\rm e} = 10^4$~K).

The density distributions in the three line components are distinctly different. We begin by discussing the densities in C1 (Fig.~\ref{fig:elecdens_C1}). Within our limited spatial coverage, there are two peaks in $n_{\rm e}$, one in the northern corner of position 1 and one in the northern corner of position 2. Position 1 covers the eastern half of region A; the peak in $n_{\rm e}$ at $\sim$2000--3000~\cmt\ is coincident with the region around cluster M82-A1 and the well-known major axis dust lane that bisects the whole nuclear region \citep[see][]{oconnell95}. From here, $n_{\rm e}$ falls off rapidly to the south and west, meaning that in the core and western half of region A (position 3) the densities are significantly lower. The peak in position 2 is again located to the north of the complex core (region C), and reaches $\sim$2000--3000~\cmt. Following the same trend, the densities then fall off towards the south, but remain fairly constant at $\sim$1000~\cmt\ throughout the rest of region C. These values are consistent with previous studies \citepalias[e.g.][]{westm07c}. The densities in positions 3, 4 and 5 fall off consistently towards the south-east (along the minor axis direction), almost reaching the low density limit ($\sim$100~\cmt) at the edge of our coverage. Interestingly, the density morphology in both the wind positions (4 and 5) follows similar linear minor axis-oriented patterns as observed previously in the FWHM and radial velocity maps \citepalias{westm09a}. In position 5, we associated these with a discrete outflow channel, but finding these type of patterns also in position 4 hints at the presence of multiple similar channels throughout the inner wind.

In C2, $n_{\rm e}$ peaks at $\sim$3000~\cmt, in the south-west of position 1 as it overlaps with position 3 (i.e.\ the core of region A). This is distinctly different to what we see in C1: the western half of region A is much denser in C2, whereas the eastern half is considerably less dense. This type of result immediately highlights the advantage of spatially resolved observations over long-slit data \citepalias[e.g.][]{westm07c}. A secondary peak in C2 densities of just over 1000~\cmt\ is seen in the centre of region C (position 2). Again the morphology of C2 densities in region C differs significantly to those in C1. The C2 densities in the wind positions (4 and 5) follow the same pattern as seen in C1 (falling off along the minor axis), but remain much higher for the same radii. The gas emitting C2 must therefore remain denser further out into the wind. It is worth noting here that the linear minor axis-oriented patterns are still visible, but appear to anti-correlate with those seen in C1. This is discussed further Section~\ref{sect:channel}.

The C3 density maps, tracing the density of the fainter component of any split lines, are noisy and very incomplete. There are, however, a number of points worth drawing attention to. C3 densities peak in the core of region C (at $\gtrsim$2000~\cmt, but with large uncertainties), approximately coincident with the flux peak. The gas knot in the centre of position 3 shows up clearly with densities between a few 100 and 1000~\cmt. And in positions 4 and 5, C3 is detected in a number of north-west--south-east strips with densities of a few 100~\cmt. This distribution, where the densities of the front- and back-sides of the expanding structures are clearly different, emphasises that the structure of M82 is highly complex.

To summarise this section, the C1 densities do not follow the H$\alpha$ emission morphology, but peak in the north of our spatial coverage, to the north of complexes A and C. This is coincident with the location of the major axis dust lane separating complexes A and C from complexes D and E \citepalias[see fig.~1,][]{westm09a}. The C2 densities are much more consistent with the H$\alpha$ flux distribution (i.e.\ the C2 densities peak within the SB complexes), but remain much higher further out into the wind compared to C1. Since C3 represents the weaker component of any split lines detected, it is interesting that its density distribution does not closely follow that of C1. Uncertainties on the C3 densities are, however, much higher than those for C1 or C2.

\subsection{Excitation diagnostics} \label{sect:GMOS_diag}
\begin{figure*}
\centering
\begin{minipage}{0.49\textwidth}
\includegraphics[width=1\textwidth]{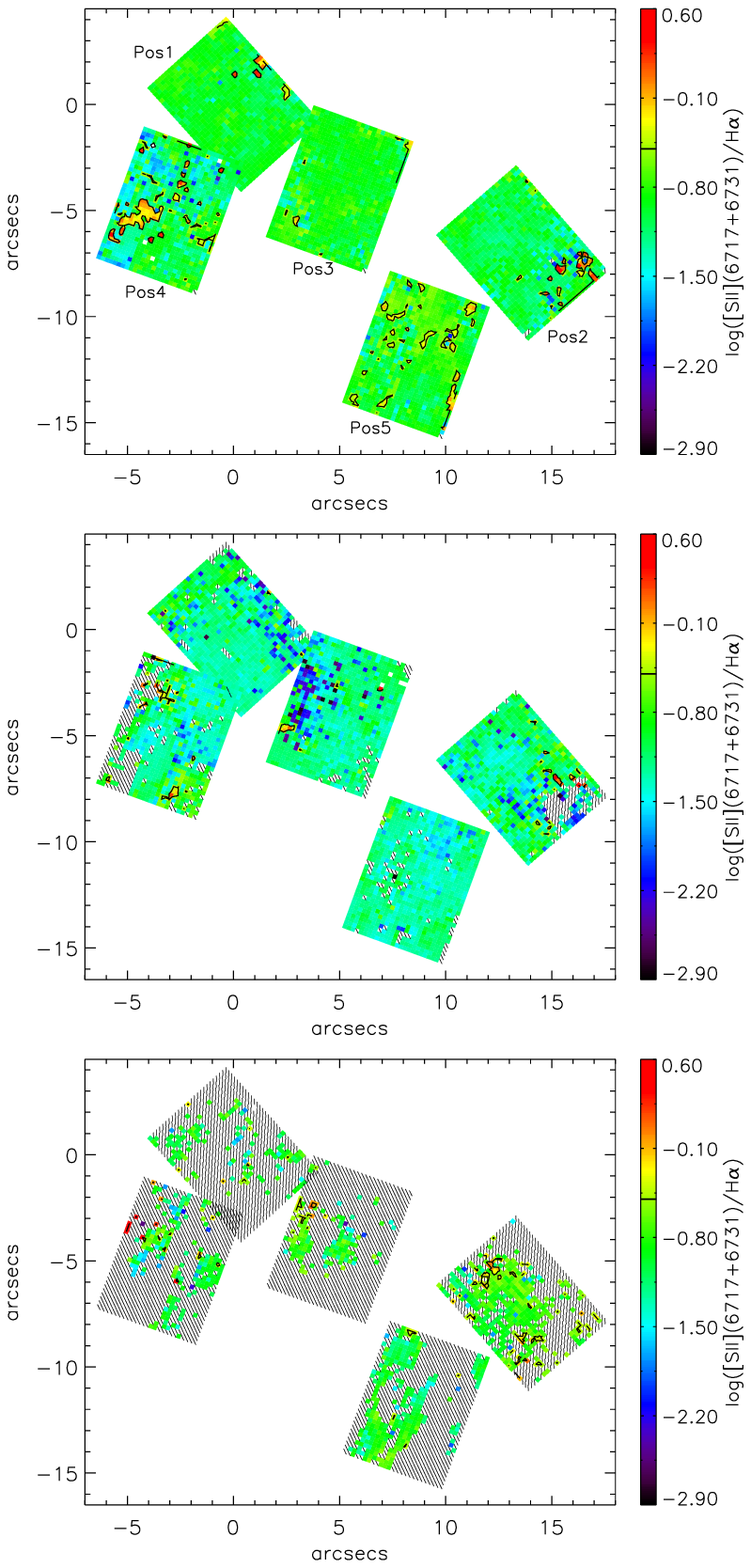} 
\end{minipage}
\begin{minipage}{0.49\textwidth}
\includegraphics[width=1\textwidth]{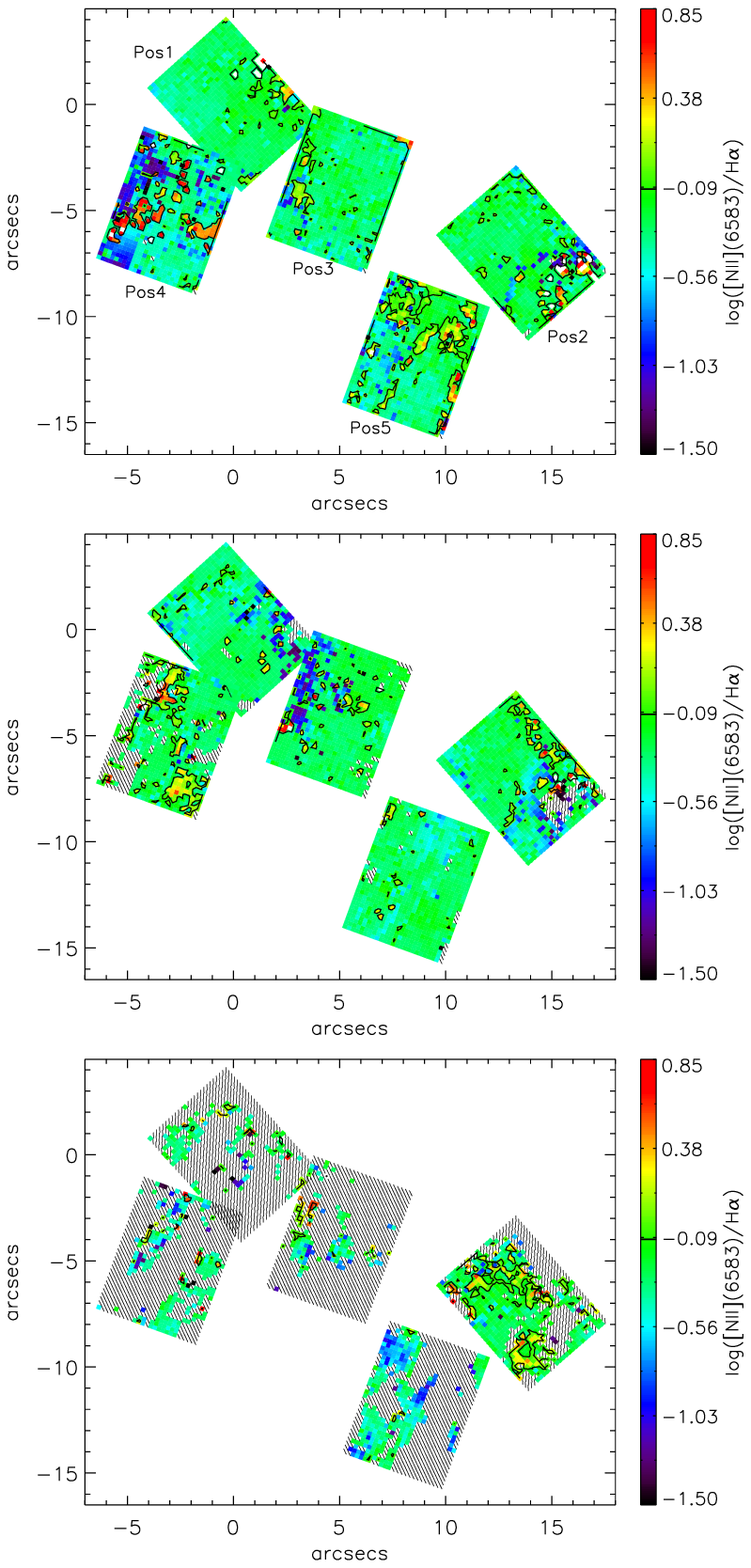} 
\end{minipage}
\caption{Maps of the flux ratio of log([S\two]($\lambda$6717+$\lambda$6731)/H$\alpha$) (left panels) and log([N\two]$\lambda$6583/H$\alpha$) (right panels). Contour lines are plotted at $-0.5$ ([S\two]/H$\alpha$ maps, as indicated in the colour-bar) and $-0.1$ ([N\two]/H$\alpha$ maps, also indicated in the colour-bar), representing fiducial ratios above which the excitation is likely to be dominated by non-photoionizing processes (see text). \textit{(A color version of this figure is available in the online journal.)} }
\label{fig:SII_NII_Ha}
\end{figure*}

The forbidden/recomination line flux ratio of [S\two]($\lambda$6717+$\lambda$6731)/H$\alpha$ can be used as an indicator of the number of ionizations per unit volume, and thus the ionization parameter, $U$ \citep{veilleux87,dopita00,dopita06b}. $U$ is nearly constant within a single Str\"{o}mgren sphere, a property that appears to hold over large scales in M82. [S\two]/H$\alpha$ is also particularly sensitive to shock ionization because relatively cool, high-density regions form behind shock fronts which emit strongly in [S\two] thus producing an enhancement in [S\two]/H$\alpha$ \citep{dopita97, oey00}. Maps of this ratio in the three line components are shown in Fig.~\ref{fig:SII_NII_Ha} (left panels), where a single contour line at log[S\two]/H$\alpha$ = $-0.5$ represents a fiducial threshold above which non-photoionized emission is thought to play a dominant role \citep{dopita95, kewley01, calzetti04}. In environments such as this, the most likely excitation mechanism after photoionization is that of shocks.

\begin{figure*}
\centering
\includegraphics[width=\textwidth]{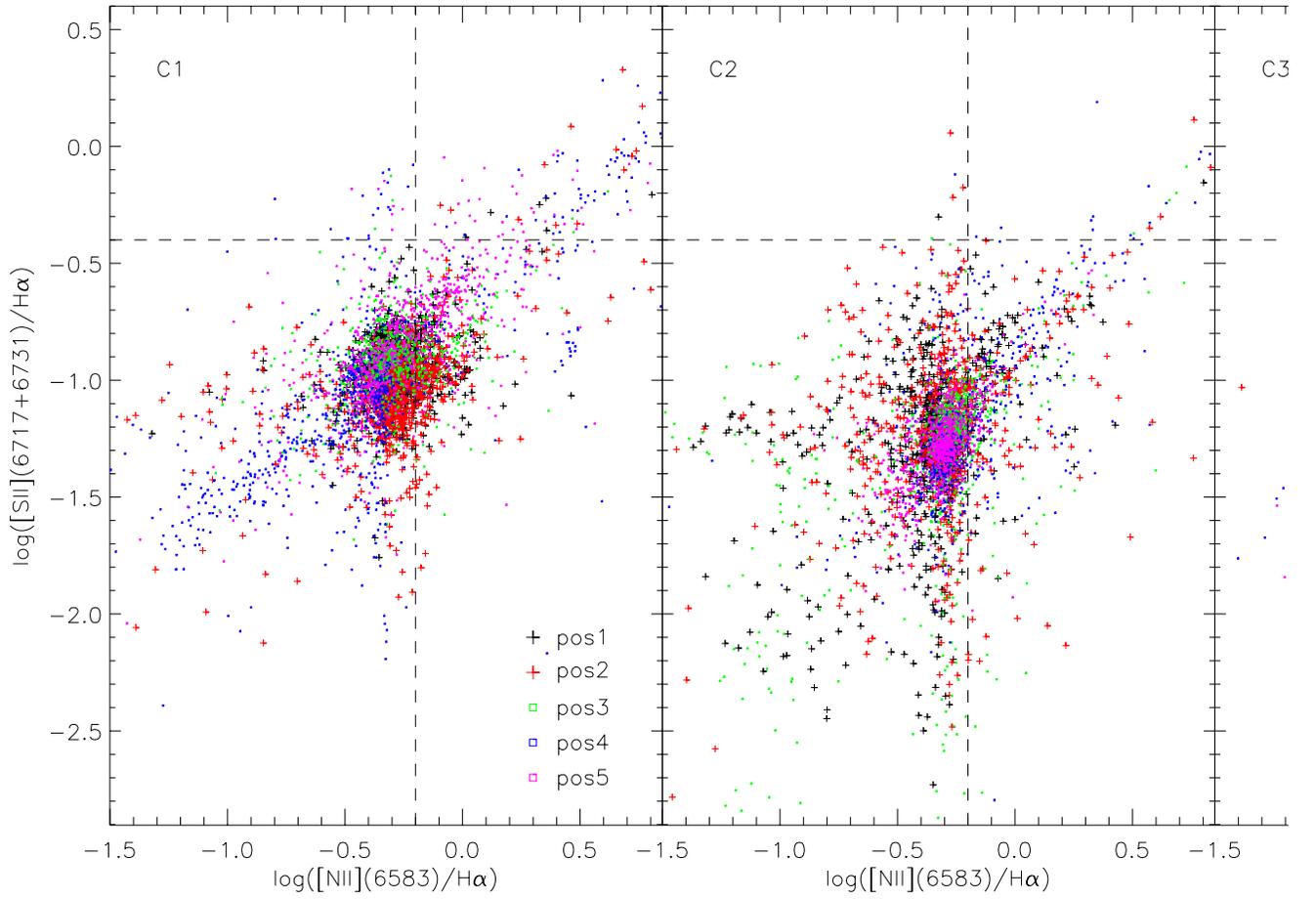} 
\caption{Plots of log([S\two]($\lambda$6717+$\lambda$6731)/H$\alpha$) vs.\ log([N\two]$\lambda$6583/H$\alpha$) for each IFU position, separated by line component. The vertical dashed lines represent the non-photoionziation thresholds from Fig.~\ref{fig:SII_NII_Ha}.
\textit{(A color version of this figure is available in the online journal.)} }
\label{fig:diagnostic}
\end{figure*}

The ratio of [N\two]$\lambda$6583/H$\alpha$ can also be used in a similar fashion to [S\two]/H$\alpha$, as a diagnostic of the gas excitation level and to search for the presence of shocks \citep{veilleux87, dopita95}. In Fig.~\ref{fig:SII_NII_Ha} (right panels), we show maps of this ratio for all three components; here the contour level represents a fiducial non-photoionized threshold of log[N\two]/H$\alpha$ = $-0.1$. These maps are qualitatively very similar to those of [S\two]/H$\alpha$, and comparing the two figures gives an idea of the significance of the variations and the uncertainties. To aid with this further, we have plotted the [S\two]/H$\alpha$ and [N\two]/H$\alpha$ points for each IFU position together in Fig.~\ref{fig:diagnostic}, separated into three graphs by line component. The dashed lines represent the aforementioned non-photoionization thresholds.

The gas excitation in C1 (Fig.~\ref{fig:SII_NII_Ha}, top panels) remains very uniform throughout the majority of positions 1, 2 and 3 (regions A and C), with mean values of log([S\two]/H$\alpha$)~=~$-0.96\pm 0.22$, and log([N\two]$\lambda$6583/H$\alpha$)~=~$-0.28\pm 0.23$. The errors quoted are the standard deviations of the samples, and their similarity shows just how constant these two ratios remain over these two positions. These results, in terms of both the absolute values and constancy, are in good agreement with previous studies (\citealt{shopbell98, forster01, smith06}; \citetalias{westm07c}).

The C2 [S\two]/H$\alpha$ and [N\two]/H$\alpha$ line ratios are both consistently lower than the average in two regions: complex A and the western half of complex C. For the most part these low excitation regions are coincident with the corresponding C2 density peaks (Fig.~\ref{fig:elecdens_C2}), suggesting that local changes in the ionization parameter play a strong part in setting the line ratios.

In components C1 and C2, high (non-photoionized) line ratios (both in [S\two]/H$\alpha$ and [N\two]/H$\alpha$) are predominantly found in positions 4 and 5. This may indicate that, as the distance from the starburst complexes increases, shock excitation begins to play an increasing role in ionizing the gas. The phenomenon of increasing line ratios with radius has been found in many starbursting galaxies on both the small and global scales \citep{heckman95, martin97, calzetti99, calzetti04}. Where the line ratios are within the photoionized regime, the radial gradient can be explained as a dilution of the radiation field from the centralised source, but in many cases shock excitation is required to explain the ratios observed that are outside the reach of pure photoionization.

Notably, one of the regions of high line ratios seen in the GMOS data, identifiable in both the C2 maps, is coincident with the `position 4 knot' identified in \citetalias{westm09a}, and discussed further in Section~\ref{sect:disc_knot}. The C1 [N\two]/H$\alpha$ map also exhibits very low ratios (log([N\two]/H$\alpha$) $<$ $-0.6$) in position 4 that are not coincident with any obvious features in the corresponding density map (Fig.~\ref{fig:elecdens_C1}). In C3, high line ratios are found to the centre-east of position 3 and to the north and east of the region C flux peak in position 2. 

In summary, the unprecedented spatial-resolution of these line ratio maps shows that (1) the excitation of the gas within the starburst complexes is extremely uniform (and consistent with pure photoionization), \textit{despite} the dramatic density gradients; and (2) regions of both high- and low-excitation gas are found within the inner wind flow, but only in very localised, distinct pockets. In C1, these regions do not correspond to anything seen in any of the other maps presented here or in \citetalias{westm09a}, but the low-excitation regions in C2 are coincident with the density peaks.

\section{Nebular properties of the disk} \label{sect:DP}

In this section we present and describe the 2D maps of H$\alpha$ line flux and derived nebular properties for the DensePak dataset. In contrast to the small-scale, high-resolution GMOS-IFU observations of the nuclear regions, these data cover the majority of the whole central disk of M82 at considerably lower spatial resolution. However, as highlighted in \citetalias{westm09a}, the spectral resolution and S/N are comparable, thus allowing us to make meaningful comparisons between the small- and large-scales as sampled by the two datasets.

\begin{table}
\begin{center}
\caption{Mean H$\alpha$ component flux ratios for the GMOS and DensePak datasets.}
\label{tbl:Ha_ratios}
\begin{tabular}{c c c}
\hline
Ratio & GMOS average & DensePak average \\
\hline
C2/C1 & 1.65$^{1}$ & 0.99 \\
C3/C1 & 0.41 & 0.43 \\
\hline
\end{tabular}

\begin{tabular}{p{8cm}}

$^{1}$average excluding ratios $>$10 since these are likely to be spurious (see text).
\end{tabular}
\end{center}
\end{table}

\subsection{H$\alpha$ maps} \label{sect:DP_Ha}
\begin{figure*}
\centering
\includegraphics[width=0.95\textwidth]{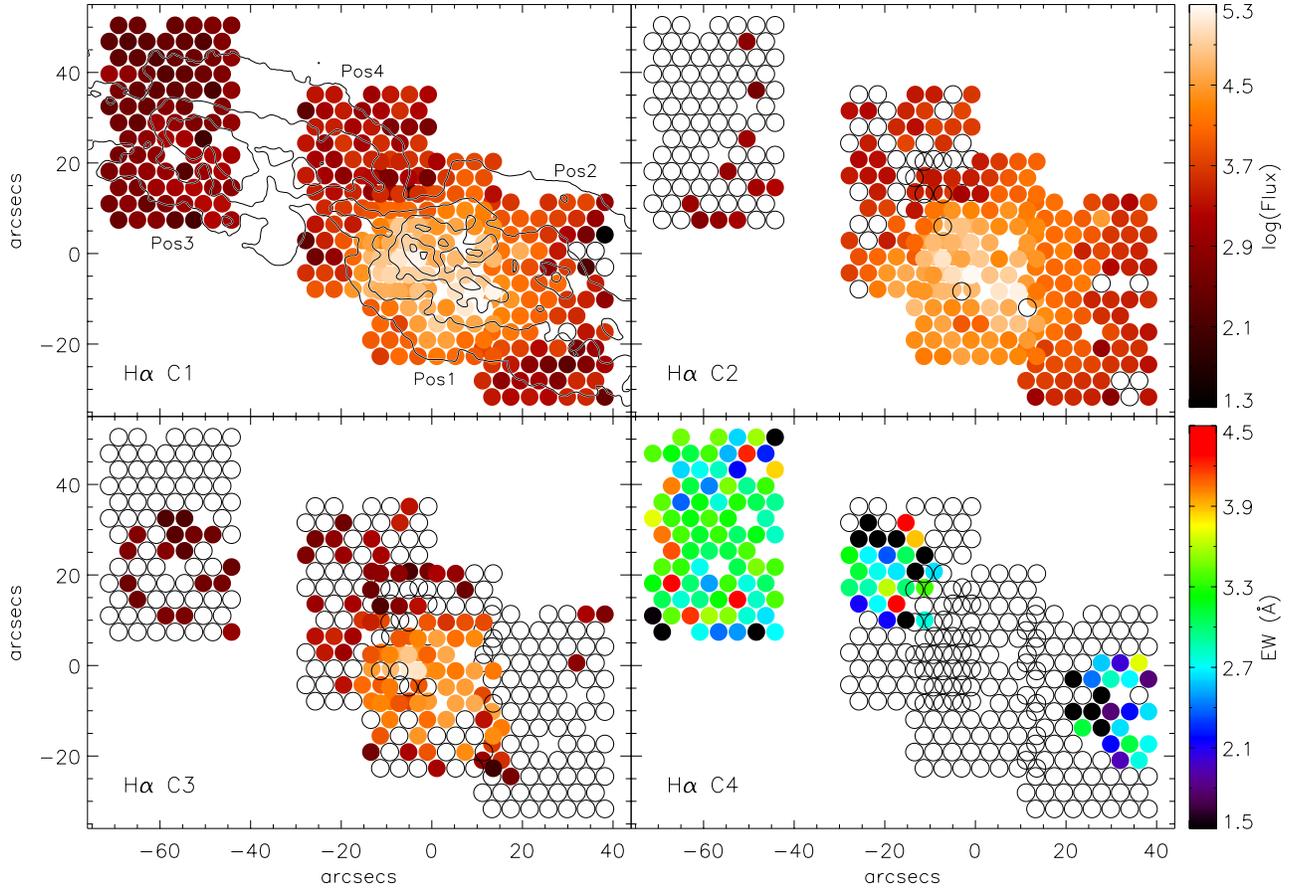} 
\caption{Line flux maps for the four identified H$\alpha$ components. The C1, C2 and C3 maps are in arbitrary (but relative) flux units, whereas the C4 (the absorption component) map is in units of equivalent width (EW) in \AA. Overplotted on the C1 map are contours from the log scaled WIYN $R$-band image (see Paper~I) to indicate the location of the starburst complexes. The origin is set to the position of the 2.2~$\mu$m nucleus. \textit{(A color version of this figure is available in the online journal.)} }
\label{fig:dp_flux}
\end{figure*}

\begin{figure}
\centering
\includegraphics[width=0.7\textwidth]{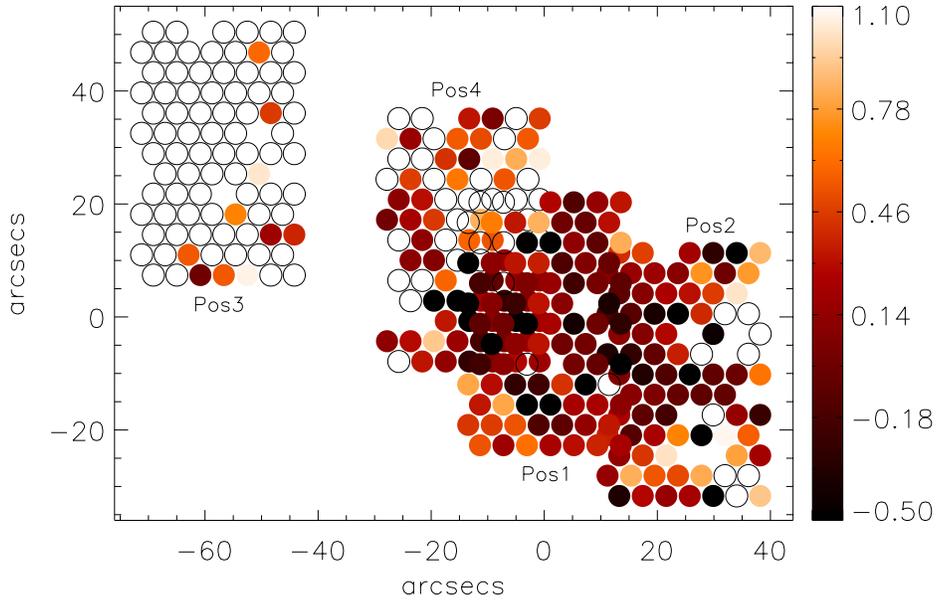} 
\caption{DensePak H$\alpha$ C2/C1 flux ratio map (in logarithmic units). As in Fig.~\ref{fig:Hac2c1_ratio}, the marker in the scale bar indicates a ratio of 1 (i.e.\ C2 = C1). \textit{(A color version of this figure is available in the online journal.)} }
\label{fig:dp_C2C1_ratio}
\end{figure}

Fig.~\ref{fig:dp_flux} shows the distribution of H$\alpha$ flux in components 1, 2 and 3, and the equivalent width (EW; in \AA) of component 4, together with contours representing the WIYN $R$-band image presented in Paper~I. As expected, the flux map of C1 (the bright narrow component) closely follows what is found through H$\alpha$ narrow-band imaging: the bright nuclear starburst complexes stand out in position 1, whereas region B to the east of the disk (position 3) is conspicuously faint in H$\alpha$. The broad component, C2, is detected over parts of all the positions, but is brightest in regions A and C and only detected in a few spaxels in region B. Over all the regions where C2 is detected it is, on average, of equal brightness to C1 (see Table~\ref{tbl:Ha_ratios}). However, the C2/C1 flux ratio is not constant across the face of the disk, as can be seen in Fig.~\ref{fig:dp_C2C1_ratio}, where we plot a map of the C2/C1 flux ratio. With this logarithmic colour stretch, it is clear that the ratio begins at low values near the disk midplane (0.3--0.5) and increases with height above and below the disk (up to $\sim$10).

Component 3 (C3), the fainter narrow component, is detected over parts of the whole central starburst area, extending out to the north of the disk (position 4) and region B (position 3). As in the GMOS data, its intensity is on average $\sim$40\% that of C1 (Table~\ref{tbl:Ha_ratios}). C4, the absorption component, is observed over the whole of region B (in position 4 and a part of position 3) and to the south-west of cluster F in position 2. An absorption component is expected in these regions since they are located outside the current starburst \citep{smith07} and hence contain an older stellar population dominated by A-type stars \citep{degrijs01, forster03, konstantopoulos08}. The H$\alpha$ flux absorbed is at a maximum in region B, where EWs of $\sim$3--4~\AA\ are seen. An example line profile from region B is shown in fig.~5d of Paper~I, clearly illustrating a broad and well-fitted absorption component.

\subsection{Electron densities} \label{sect:DP_elecdens}
\begin{figure*}
\centering
\includegraphics[width=\textwidth]{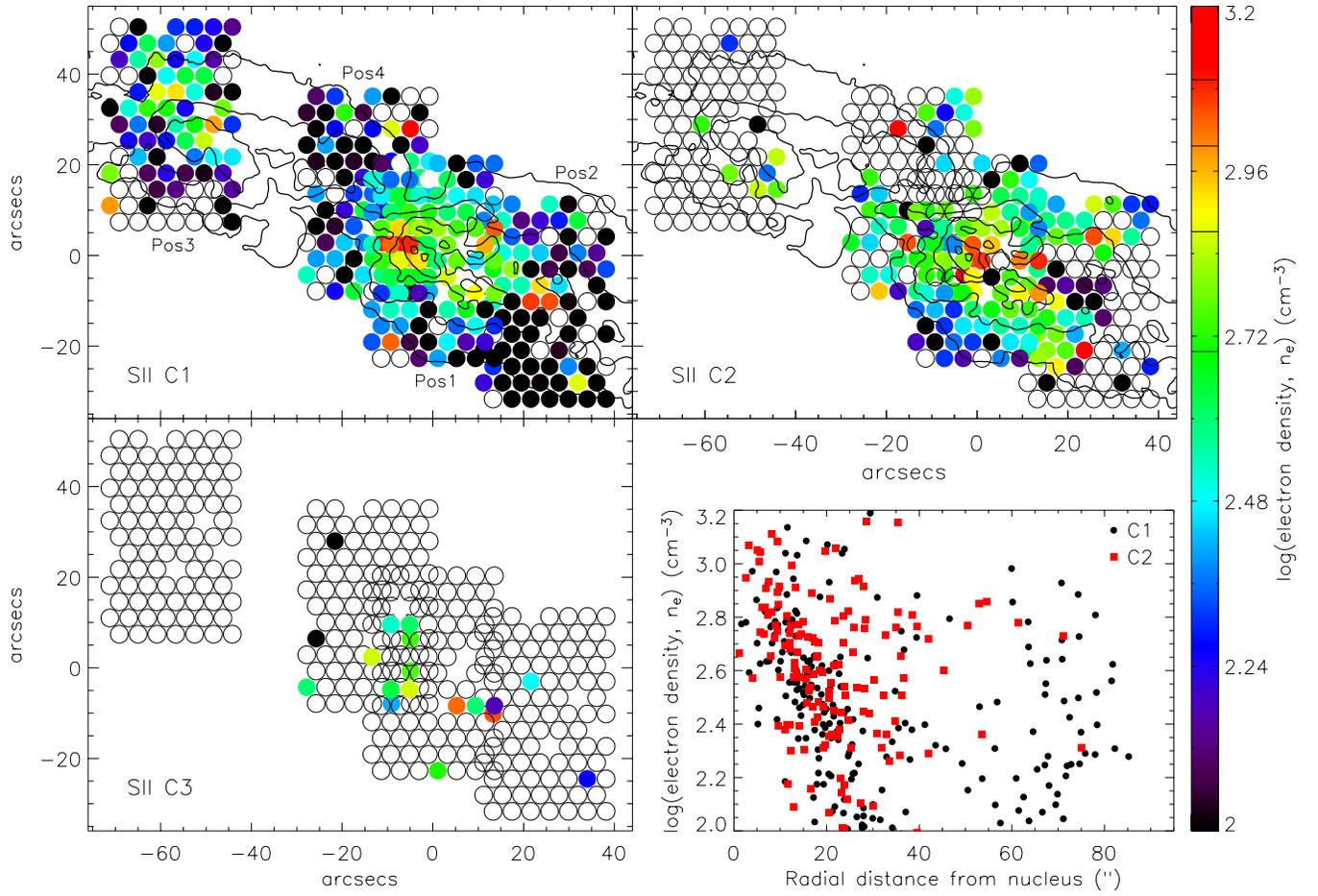} 
\caption{Maps of the [S\two] derived electron density, $n_{\rm e}$, in the four DensePak fields. Units are in \cmt; markers in the scale bar indicate 500, 750, 1000 and 1250~\cmt. Uncertainties lie in the range $\pm$500 for the highest densities to $\pm$50 for the lowest. Overplotted on the C1 and C2 maps are contours from the log scaled WIYN $R$-band image (see Paper~I) to indicate the location of the starburst complexes. The lower-right-hand panel shows a radial plot of the electron densities of C1 and C2, highlighting the more extended distribution of higher C2 densities compared to C1. \textit{(A color version of this figure is available in the online journal.)} }
\label{fig:dp_elec_dens}
\end{figure*}

Maps of the electron densities, $n_{\rm e}$, measured from the [S\two] ratio (assuming $T_{\rm e}=10^{4}$) are shown in Fig.~\ref{fig:dp_elec_dens}. Note, the velocities and FWHMs of the three [S\two] components are consistent with those of H$\alpha$ (Paper~I) meaning that the components identified in both lines physically correspond to one another, despite the lower S/N of the [S\two] line. In C1 the highest densities are found in the eastern half of region A, peaking at $\sim$1000~cm$^{-3}$. The location of this peak agrees well with our GMOS data; its somewhat lower value can be attributed to the difference in spatial resolutions between the two datasets (i.e.\ the much finer spatial resolution of GMOS allows us to pick up the small-scale, localised regions of higher density gas). The peak density in region C is lower \citepalias[at 500--700~\cmt; in good agreement with][]{westm07c}, and densities remain at a level of a few 100 throughout the central starburst region, particularly on the northern side. Another peak is found in region B, with densities reaching 500--700~\cmt.

In C2 densities rise to $\gtrsim$1000~\cmt\ in the spaxels covering the western half of region A and region E, in excellent locational agreement  with our GMOS data (again the lower absolute values can be attributed to the coarser spatial sampling of DensePak). Densities remain higher to much larger radii in C2, both in the major and minor axis directions. In order to highlight this further, we have plotted the radial distribution (relative to the nucleus) of both C1 and C2 electron densities in the lower-right-hand panel of Fig.~\ref{fig:dp_elec_dens}. This clearly shows that the C2 densities remain much higher out to larger radii relative to C1. This intriguing finding extends the trend we found in the GMOS data to the entire disk, and is discussed in relation to the origin of C2 in Section~\ref{sect:disc}.

\citet{oconnell78} were the first to recognise that ionized gas densities in the starburst core of M82 were unusually high, finding [S\two]-derived electron densities of $\sim$1800~\cmt\ in regions A and C, and subsequent multi-angle long-slit observations showed that the gas densities rapidly fall off radially \citep{heckman90}. The minor axis decrease was accurately measured by \citet{mckeith95} using deeper long-slit data. They found electron densities of $\gtrsim$1000~\cmt\ near the nucleus (where they only detect their strongest line component), and by $\pm$500~pc the densities had fallen to $\sim$100--200~\cmt. \citetalias{westm07c} presented \textit{HST}/STIS spectroscopy of the starburst core and the first evidence for how rapidly the densities vary with position: over only 2.5~arcsecs (45~pc) they found the densities to fall from $\sim$1800~\cmt\ \citep[near cluster A1; see also][]{smith06} to $\sim$1000~\cmt\ in the centre of region A and 600--800~\cmt\ in region C. 

Our DensePak measurements thus agree very well with these studies. The coarse spatial resolution of DensePak explains why we do not see the highest densities mentioned in the literature and seen with our GMOS data. Having said that, our DensePak observations have the advantage of both of both wider two dimensional spatial coverage and high S/N, meaning we have been able track the density of both C1 and C2 over a significant proportion of the inner disk. Finally, overall the densities of both C1 and C2 stay higher on the northern side of the disk compared to that of the southern side. This can be understood as being a result of the 80$^{\circ}$ inclination of the disk, such that the northern side is nearest to the observer \citepalias[see fig.~13 of][]{westm07c}.

The main finding of this section is that although the the C1 and C2 densities peak at approximately the same value in the centre of the starburst, the C2 densities remain higher to much larger projected radii (i.e.\ both in the major and minor axis directions) than their C1 equivalents.

\subsection{Excitation diagnostics} \label{sect:DP_diag}

\begin{figure}
\centering
\includegraphics[width=\textwidth]{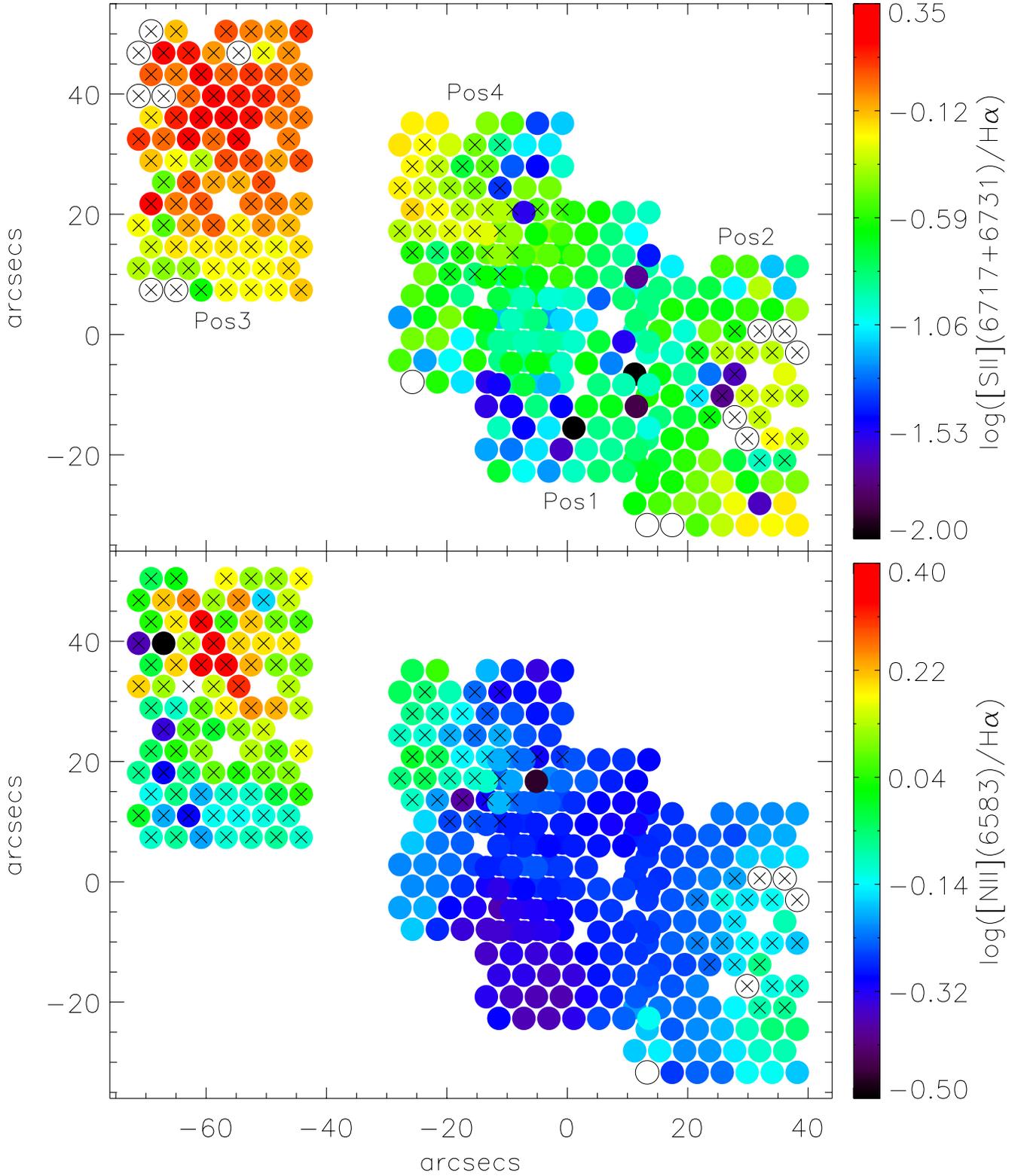} 
\caption{Flux ratio maps of log([S\two]($\lambda$6717+$\lambda$6731)/H$\alpha$) (upper panel) and log([N\two]$\lambda$6583/H$\alpha$) (lower panel) -- note the different colour bar scales. The ratios were calculated after summing the flux from all identified components of the lines. Spaxels marked with crosses are those that contain an identified H$\alpha$ absorption component, illustrating how the absorption is biassing the line ratios to artificially high values. \textit{(A color version of this figure is available in the online journal.)} }
\label{fig:dp_lineratios}
\end{figure}

Fig.~\ref{fig:dp_lineratios} shows maps of the [S\two]($\lambda$6717+$\lambda$6731)/H$\alpha$ and [N\two]$\lambda$6583/H$\alpha$ line ratios. Since not all line components were detected in each line, we have summed the flux of all identified components for each line before calculating the line ratios.

A significant range in values are seen in both maps, with elevated ratios in region B (position 3 and the north east of position 4) and to the south-west of cluster F (position 2). However, a closer inspection of the spectra from these regions reveals that the majority of spaxels with these high ratios exhibit an H$\alpha$ absorption component -- we have marked spaxels where an absorption component is detected with a cross in order to highlight the correspondence. Since absorption is only present in recombination lines, the forbidden/recombination line ratios become biassed towards artificially high values when the absorption is present. This highlights the need to be careful of absorption effects in line ratios, particularly for ratios derived from images, or simple single Gaussian fitting, where absorption cannot be accounted for.

Disregarding these H$\alpha$-absorption affected regions, the two line ratios remain fairly uniform (particularly [N\two]/H$\alpha$). The mean ratio of log([N\two]/H$\alpha$) for all spaxels not containing absorption is $-0.3\pm 0.1$, and similarly for log([S\two]/H$\alpha$) is $-0.8\pm 0.3$. These values are in very good agreement with the GMOS data (Section~\ref{sect:GMOS_diag}).

\citet{oconnell78} find a log([N\two]/H$\alpha$) ratio for the nuclear regions (including complexes A, C and E) of $-0.28$ from long-slit data. \citet{shopbell98} find uniform and low values of log([N\two]/H$\alpha$) in the inner 1~kpc of the M82 outflow ranging from $-0.5$ to $-0.2$. Both these results are consistent with our measurements, and in very good agreement with \citetalias{westm07c}. The line ratio maps of \citet{shopbell98} also show very high values further out in the disk (to the east and west of the nucleus), in the same place as we find exaggerated values due to H$\alpha$ absorption. At minor axis heights of 1~kpc or more ($>$$60''$), \citet{shopbell98} find increasing values of [N\two]/H$\alpha$ which they attribute to a dilution of the radiation field in the outer-wind together with a possible increasing influence of shocks. However, our data do not reach this far in radial extent so we do not see this effect.

\section{Discussion} \label{sect:disc}

Using all the information we now have about the gas conditions within the M82 starburst, we can begin to put together a model of the ISM in the starburst core, and examine how these conditions change as we move out of the disk into the wind.

\subsection{ISM conditions within the starburst core} \label{sect:disc_disk}

The electron densities we have mapped peak in the starburst core, fall off dramatically as we leave the disk, and exhibit significant local variations. The peak values (few 1000~\cmt) imply very large interstellar pressures (P/k $>$ $10^7$). Although these results have been found before by a number of independent studies (e.g.\ \citealt{heckman90, mckeith95}; \citetalias{westm07c}), our high-resolution, two-dimensional maps show that the peaks are very localised and that there is a large amount of structure within the density distribution in the starburst core.

Contrary to expectation, the C1 densities peak not in the starburst complexes themselves, but to the north, in the major axis dust lane. Recalling one of our conclusions from \citetalias{westm09a} -- that the emission of C1 is weighted towards the inner-most regions of the starburst (i.e.\ the base of the outflow) -- we see that these density results support this idea, and indicate that the body of bright, \textit{relatively} quiescent gas from which C1 is emitted, is not associated directly with the main starburst complexes themselves, but extends well beyond the complexes into the dense dust lane. C2, however, is at its densest within complex A (and to a lesser extent C); its overall distribution follows the flux morphology much more closely than C1. Its peaks are also of a few 1000~\cmt\ implying large pressures. These results are again in support of our conclusions from \citetalias{westm09a}, where we suggested that the C2 emission is weighted towards regions further out along the line-of-sight.

In this paper we have also presented evidence at high spatial resolution for the impressive uniformity of the line ratios and excitations within complexes A and C (particularly in component C1), first highlighted by studies such as \citet{forster01} and \citet{beirao08}. We have also confirmed how this uniformity can be extended to the whole inner disk region covered by the DensePak fields-of-view. The unprecedented spatial-resolution and spatial coverage of the line ratio maps shows that (1) the excitation of the gas within the starburst complexes is extremely uniform (and consistent with pure photoionization), \textit{despite} the dramatic density gradients; and (2) regions of both high- and low-excitation gas are found within the inner wind flow, but only in very localised, distinct pockets.

We will now use these density and excitation results, together with work from the literature, to build up a model to describe the ISM conditions within the starburst core. As described in the introduction, the modelling of observations of gas in the central regions made from the opticalÐradio indicate a fragmented (shredded/fractious) ISM, consisting of a range of clouds from small/dense clumps with low filling factors ($<$1~pc, $n_{\rm e}$\,$\gtrsim$\,$10^4$~\cmt) to larger filling factor, less dense gas \citep[e.g.][]{lord96, forster01, fuente08}. These are bathed with an intense radiation field and embedded in an extensive high temperature ($T$\,$\gtrsim$\,$10^6$~K), X-ray-emitting ISM. Emission measure results from \citetalias{westm07c} suggest that the most compact clouds are found in the clump cores where the star formation is most intense and the gas pressures are high, whereas the cloud sizes in the inter-complex regions are larger. The observed near-constant state of the starburst (ionization parameter and line diagnostics) must therefore be a consequence of the small cloud sizes, allowing the gas conditions to respond quickly to any changes. This is rather akin to the thermalisation of a large number of individual particles in a plasma. To this picture, we are now able to add that the denser clouds are not only bathed in UV but are also interacting with an extensive, very hot, and presumably turbulent ISM phase that is responsible for the X-ray emission. The left-centre panel of Fig.~\ref{fig:super_fig} contains a schematic representation illustrating these ideas. Since we associate the broad component C2 with emission from turbulent mixing layers on the surface of interstellar clouds, the existence of so many small, compact, high density clumps embedded in a hot, fast-flowing wind must make for an ideal environment for the copious emission of broad lines. These surface wind-clump interaction layers must also be where mass is evaporated and stripped off the clouds by the intense radiation field and wind flows. This suggests that the ma jority of the mass loaded into the wind must originate in these clumps, where the density of the gas being stripped is highest. This argues strongly against C2 originating from scattered light from the nucleus. A further implication of the existence of so many dense clouds in complexes A and C is that there must still be a significant gas supply within the starburst core, despite the intense star formation, heavy mass-loading and wind outflows.

In our model, C2 is only emitted from turbulent mixing layers on cloud surfaces in direct line-of-sight with a wind source, whereas C1 can be emitted from any $10^4$~K gas located within the Str\"{o}mgren radius of the starburst. This includes the static ionized layers beneath the TMLs (bottom panel of Fig.~\ref{fig:super_fig}), ionized gas clouds which are not in a direct line-of-sight with an ionizing/wind source (i.e.\ wind shadowed -- ionizing radiation is easily scattered -- an example of a cloud in a wind shadow is illustrated in the bottom-left corner of the centre-left panel in Fig.~\ref{fig:super_fig}), and general diffuse ionized gas. Furthermore, the excitation of the gas within the starburst complexes is extremely uniform, and consistent with pure photoionization. This can be explained through the well-mixed nature of the clusters and clouds, which transforms the presumably very inhomogeneous distribution of ionizing photons from the clusters into a uniform escaping radiation field. A helpful analogy might be to liken the surfaces of the starburst complexes to the surface of the Sun, where there are many gas phases (temperatures and densities), a strong density gradient, chaotic dynamics and morphology, and denser pockets of gas in the form of loops and filaments (shredded shells/bubbles) entrained in the outflow/wind, all being accelerated outward. 

As a final note to this section, we turn our attention to the the $\sim$7~Myr old super star cluster M82-A1 \citep{smith06}. In that study we found M82-A1 to have followed a non-standard evolutionary path for a star cluster, since it is surrounded by a large, pressurised H\two\ nebula that is stifling its wind \citep[see also][]{silich07}. If the H\two\ shell forms a full sphere around the cluster and this stage is not simply transient, then it is difficult to explain how the superwind is powered if one assumes M82-A1 to be representative of a typical cluster in region A. However, our GMOS maps show that M82-A1 is in fact embedded in one of the highest C1 density peaks in the starburst core. Recalling that the inclination of the disk is such that on the southern side we are seeing up into the starburst \citepalias[the northern side is partially obscured by the front of the disk; see e.g.][]{westm07c}, we can now see that the rapid fall 
in densities from the north of our IFU fields into the cores of complexes A and C shows that the conditions towards the south of the complexes are much more favourable to the escape of the individual cluster winds. This implies that clusters towards the outside surfaces of the complexes may in fact be the primary drivers of the superwind.

\subsection{Inner wind} \label{sect:disk_wind}

As we move away from the plane of the disk, there are a number of results that indicate that emission from C2 becomes increasingly more important (compared to C1). Firstly the C2/C1 flux ratio increases (all the way through our GMOS fields out to the extent of our DensePak data; Sections~\ref{sect:GMOS_Ha} and \ref{sect:DP_Ha}). Since the C2/C1 ratio reflects the proportion of the volume occupied by TML vs.\ standard photoionized gas, this indicates that within the energy injection zone (thought to be defined by the minor axis velocity inflection points; \citealt{mckeith95, shopbell98}; \citetalias{westm07c}) the turbulent motions must play an increasing role with radius. Secondly, the electron densities in C2 remain higher to larger radii (Sections~\ref{sect:GMOS_elecdens} and \ref{sect:DP_elecdens}), suggesting that away from the dense core of the galaxy, then density in the TMLs is preferentially higher than in the narrow line-emitting gas. Although the majority of the mass loading might take place within the starburst complexes, the region of mass loading must extend far out into the inner wind. Thirdly, the line widths of C2, although already high in the nuclear regions, also increase with height above the disk \citepalias{westm09a}, peaking at the location of the \citet{mckeith95} minor axis velocity inflection points. Finally, it is possible that these effects are also partly due to the weakening or dilution of the C1 emission with height. What happens to these relationships beyond the energy injection zone (beyond the minor axis velocity inflection points) is something that should be investigated.

On average the C2/C1 flux ratios are considerably higher than what was found in NGC 1569 \citep{westm07b}. Since C2 is produced at the interface layers of interstellar clouds, the significantly larger gas mass and number of clusters in M82 has resulted in C2 being more prominent and may mean that a larger percentage of mass is being loaded into the outflow.

\subsection{Outflow channel} \label{sect:channel}

In \citetalias{westm09a} we identified a discrete outflow channel in IFU position 5 characterised by expanding, narrow-line-emitting walls surrounding a cavity of broad-line-emitting gas. With the additional results that we have presented here, we can explore the characteristics and implications of this outflow channel further.

\begin{figure}
\centering
\includegraphics[width=0.30\textwidth]{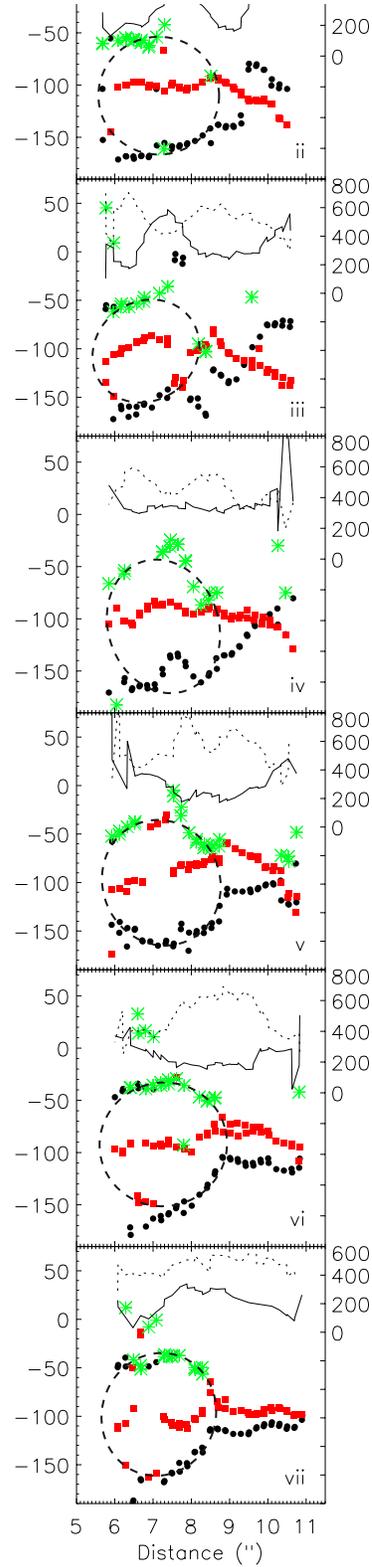} 
\caption{Position-velocity plots for seven evenly spaced (every $2''$) pseudo-slits positioned parallel to the major axis, stepping down through GMOS IFU position 5 (i-iv = north-south). H$\alpha$ C1 is plotted with black circles, C2 with red squares, and C3 with green stars. The dashed ellipses highlight the expanding outflow channel structure traced by H$\alpha$ C1 and C3; the velocity of the broad H$\alpha$ C2 indicates that it originates from within the channel. Also plotted are the corresponding C1 (solid line) and C2 (dashed line) electron densities (C3 has been excluded for clarity but is consistent with C1), showing the anti-correlation in C1 and C2 densities within the channel (C1 densest in the channel centre and falls off with radius, whereas C2 is densest in the channel walls). \textit{(A color version of this figure is available in the online journal.)} }
\label{fig:pos5_cuts_pv}
\end{figure}

\begin{figure}
\centering
\includegraphics[width=0.9\textwidth]{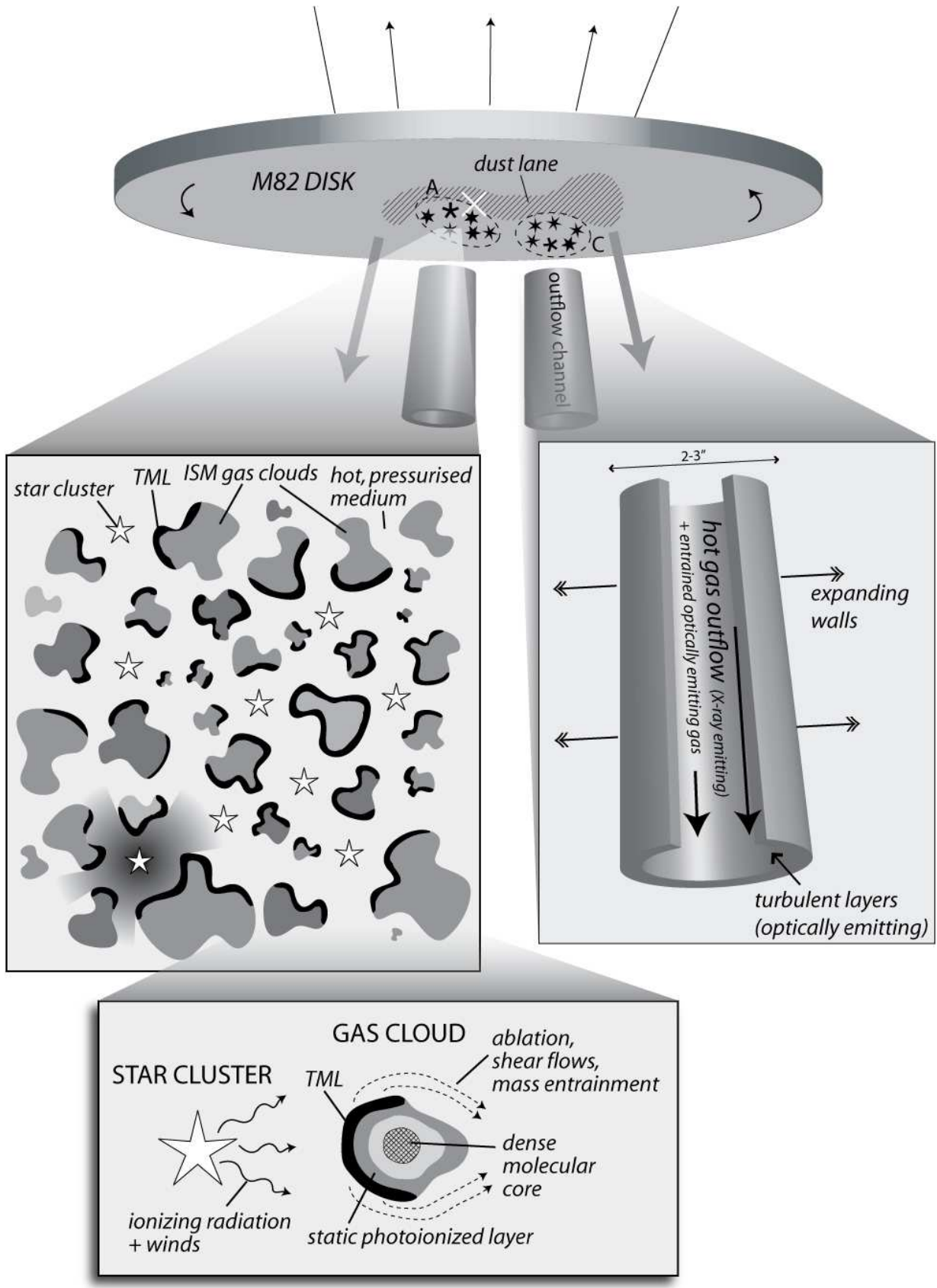} 
\caption{Cartoon model depicting the ideas developed in this work regarding the state of the ISM in M82. At the top is a representation of the disk showing the location of starburst complexes A and C, the nucleus (white cross), and the major axis dust lane (hatched region). 
The \textbf{left-hand panel} shows schematic model of the ISM in the core of one of the starburst complexes. The gas is fragmented into a large range of cloud sizes, the star clusters are well mixed in with the clouds, and both are embedded in a hot medium at very high pressure. The characteristic size-scales of the clouds and the clusters are similar (few pc). Turbulent mixing layers (TMLs) form on the surfaces of the clouds directly facing the oncoming winds from the clusters, whereas the surfaces of all the clouds are fully ionized by the energy supplied by the star clusters, even if they are located in ``shadowed'' regions (an example of a cloud in a wind shadow is illustrated at the bottom-left). 
The \textbf{bottom panel} shows a single cloud being irradiated by ionizing photons and impacted by the fast-flowing wind from a nearby star cluster. Following the idea of \citet{binette09}, a stratified cloud structure forms, with a TML on the cloud surface facing the cluster (from which the broad line is emitted), followed by a static photoionized layer (narrow lines). The core of the cloud is formed from dense molecular material. 
\textbf{Inset right:} a cartoon schematic showing the physical characteristics of the outflow channel seen in IFU position 5. The kinematics indicate a small increase in the width of the channel from top to bottom of the IFU FoV, and the walls are measured to be expanding at a roughly constant $\sim$60~\kms. Broad line emission from turbulent gas originates in the cavity of the channel, where the hot outflowing wind interacts with the cooler channel walls and other entrained material. }
\label{fig:super_fig}
\end{figure}

Fig.~\ref{fig:pos5_cuts_pv} shows H$\alpha$ position-velocity plots extracted from seven evenly spaced (every $2''$) pseudo-slits positioned parallel to the major axis, stepping down through GMOS IFU position 5. Also plotted are the C1 and C2 electron densities extracted from the same pseudo-slits (the C3 densities have been omitted for clarity, but are consistent with those of C1; see Figs.~\ref{fig:elecdens_C1} and \ref{fig:elecdens_C3}). A coherent, expanding cylinder, as indicated by the overlaid dashed ellipses, can be traced in the narrow-line (C1 and C3) kinematics extending along the whole length of position 5. These velocity ellipses show that the structure, while $>$$7''$ (120~pc) in length, increases slightly in physical width from the north to south from $\sim$2--3$''$ (35--50~pc), and maintains an approximately constant (subsonic) expansion velocity of $\sim$60~\kms. This is indicated in a schematic cartoon of the outflow channel in the right-hand panel of Fig.~\ref{fig:super_fig}. The velocities of the broad C2 line suggest that it originates within the space between the two walls, with central velocities at rest with respect to the C1-C3 expansion. The [S\two]-derived electron density plots demonstrate how the walls of the channel are defined by peaks in the C2 densities (700--1000~\cmt) and troughs in the C1 densities ($\sim$200~\cmt), whereas the centre of the channel is defined by higher C1 densities ($\gtrsim$500~\cmt) that decrease with radius. What is not shown in this plot is that the widths of C2 appear to be broader ($\sim$250~\kms) near the channel walls and narrower ($\gtrsim$150~\kms) in the centre \citepalias{westm09a}.

Thus, in our model the channel cavity contains the hot outflowing wind fluid. As it flows outwards it interacts with the cooler, denser walls of the channel and with clumps of entrained material to produce broad-line emission, while the walls themselves emit primarily the narrow lines C1 and C3. Finally, we note that the linear minor axis-oriented patterns observed in the density, FWHM and radial velocity maps (Section~\ref{sect:GMOS_elecdens}; \citetalias{westm09a}) hint at the existence of multiple similar channels within the southern outflow, as is predicted by models such as those of \citet{t-t03}.

The presence of organized channels which funnel shock-heated gas outwards from starburst complexes has a number of important implications. These structures resemble scaled-up versions of the cluster wind channels described by \citet{t-t03}, and as in their models, will tend to collimate hot gas outflows. The collimated outflows are a possible source for the well defined loop structures of ionized gas seen in H$\alpha$ emission above the main starburst complexes in M82. Indeed hints of the channel can be seen in the H$\alpha$ morphology \citepalias[see e.g.\ fig~1 in][]{westm09a} in the form of enhanced emission following the channel walls. Without the line ratio and kinematical evidence, however, it would have been impossible to distinguish from any other formation in this chaotic region. Channelled flows will also influence the mass loading process, and it seems likely that some of the gas within such columns could experience minimal mass loading. In this case they will improve the efficiency in launching a galactic wind in that some hot gas can flow out of the disk relatively unimpeded. In addition outflow channels can provide low opacity pathways for the escape of UV radiation from starburst complexes which in turn can help ionize the large quantities of extraplanar gas associated with this galaxy.

\subsection{Position 4 knot} \label{sect:disc_knot}

\begin{table*}
\begin{center}
\caption{Emission line Gaussian component measurements from the summed spectrum (Fig.~\ref{fig:knot_spec_fits}) of the position 4 knot.}
\label{tbl:knot_spec}
\begin{tabular}{c c c c c c}
\hline
Component & H$\alpha$ FWHM & H$\alpha$ radial velocity & log([N\two]/H$\alpha$) & log([S\two]/H$\alpha$) & Density \\
& (\kms) & (\kms) & & & (\cmt) \\
\hline
C1 & 156 & 40 & $-0.34$ & $-0.59$ & 480 \\
C2 & 333 & $-138$ & $-0.43$ & $-0.64$ & 1100 \\
C3 & 53 & 37 & $-0.42$ & $-0.81$ & 900 \\
\hline
\end{tabular}
\end{center}
\end{table*}

\begin{table*}
\begin{center}
\caption{Point source photometry, where $U$ = F330W, $B$ = F435W, $V$ = F555W, $I$ = F814W.}
\label{tbl:knot_photom}
\begin{tabular}{l c c c c c c c c c c c}
\hline
Object & \multicolumn{2}{c}{Coordinates} & $V$ & $M_{V}$ & $U-V$ & $B-V$ & $V-I$ & $B-I$ & $U-I$ & $U-B$ \\
& \multicolumn{2}{c}{(J2000)} & \multicolumn{8}{c}{--- (mags) ---} \\
\hline
1 (pos4 knot) & $09^{\rm h}\,55^{\rm m}\,54\fsec06$ & $69^{\circ}\,40'\,44\farcs35$ & 22.16 & $-5.22$ & 1.14 & 0.98 & 1.40 & 2.38 & 2.45 & 0.16 \\
2 & $09^{\rm h}\,55^{\rm m}\,53\fsec03$ & $69^{\circ}\,40'\,39\farcs60$ & 20.46 & $-6.92$ & 0.68 & 0.62 & 0.79 & 1.41 & 1.47 & 0.06 \\
\hline
\end{tabular}
\end{center}
\end{table*}

\begin{figure}
\centering
\includegraphics[width=0.45\textwidth]{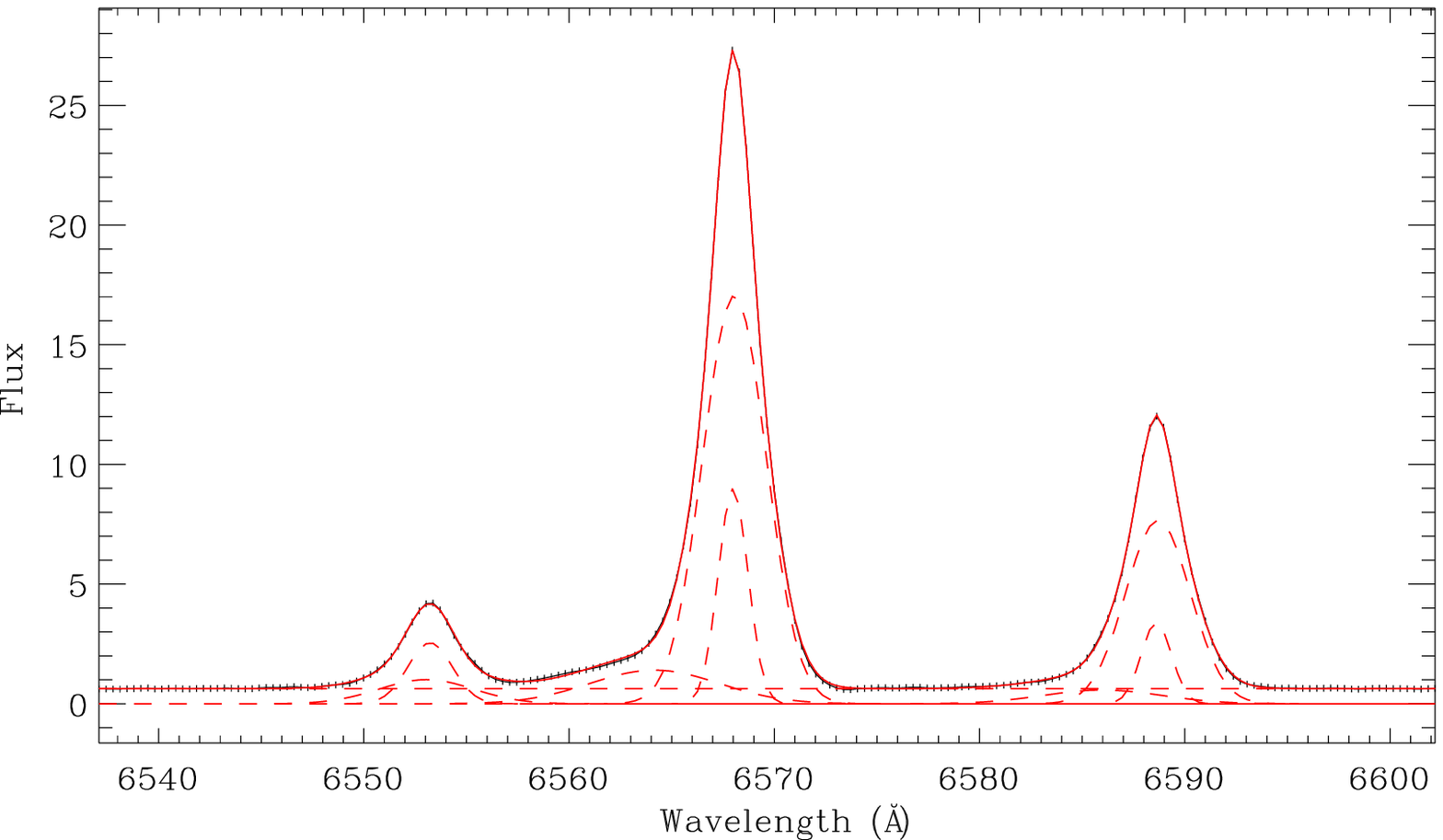} 
\includegraphics[width=0.45\textwidth]{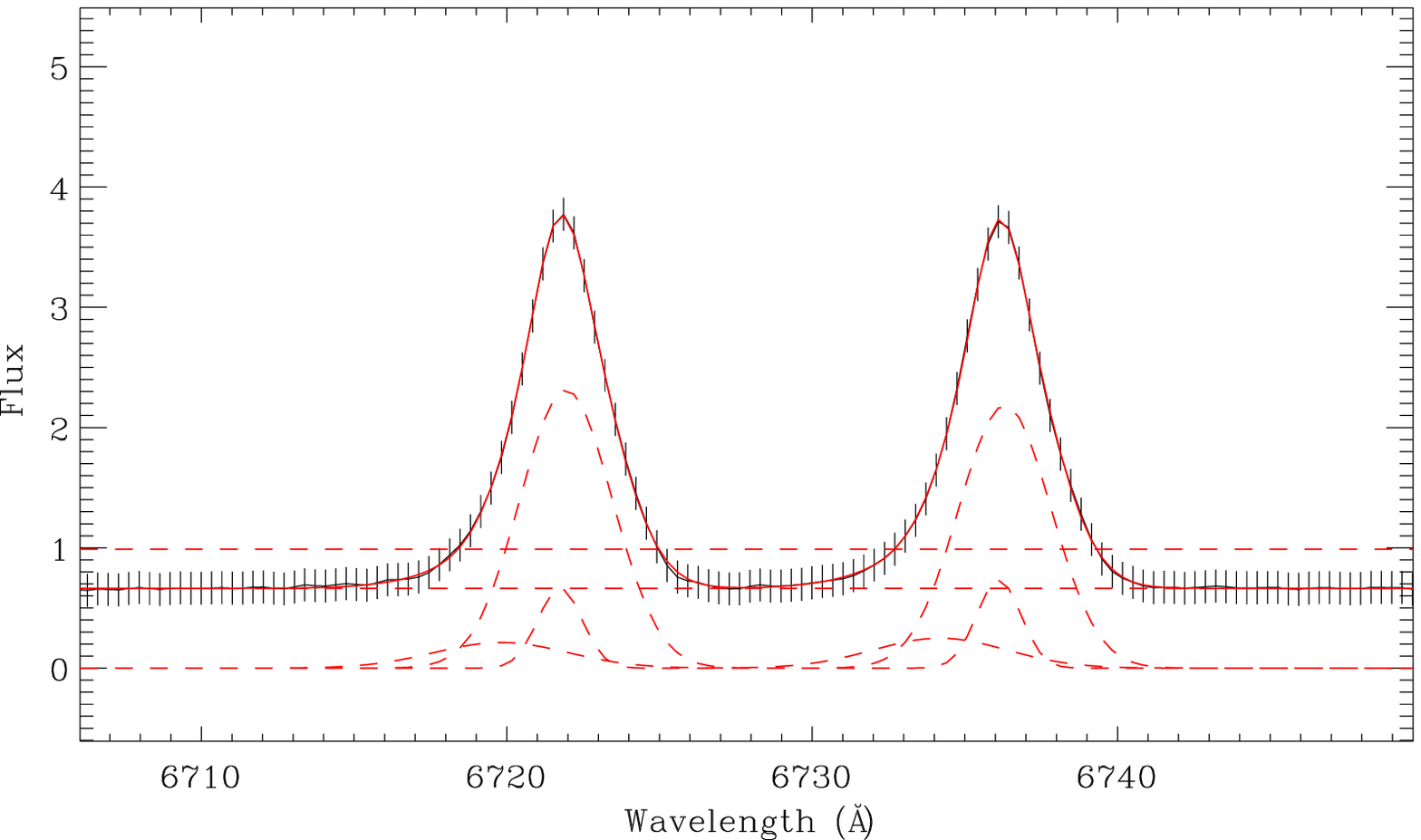} 
\caption{Summed spectrum (over 10 spaxels) of the position 4 knot, with fits to the H$\alpha$, [N\two] and [S\two] lines. \textit{(A color version of this figure is available in the online journal.)} }
\label{fig:knot_spec_fits}
\end{figure}

In Paper~I we identified a small region in GMOS IFU position 4, coincident with a bright, point-like object on \textit{HST}/ACS imaging, with distinctly blueshifted (by $\sim$140~\kms) and broad ($\lesssim$350~\kms) H$\alpha$ emission that we referred to as the ``position 4 knot''. From the data presented in this paper, we see that this knot does not show up in either the flux or density maps. However, a small number of spaxels near the location of the knot exhibit a slight enhancement in [S\two]/H$\alpha$ C2 and a significant enhancement in [N\two]/H$\alpha$ C2 compared to the surroundings (Fig~\ref{fig:SII_NII_Ha}). To confirm these results, we extracted and summed the spectra from 10 spaxels centred on the knot; the resulting spectra and multi-Gaussian line fits are shown in Fig.~\ref{fig:knot_spec_fits}. The broad (FWHM $\sim$ 330~\kms), blueshifted ($-140$~\kms) component is clearly identifiable in all five emission lines. The line component properties are given in Table~\ref{tbl:knot_spec}. The electron densities derived from the [S\two] line ratios indicate that the gas from which the broad component is being emitted is denser than the narrow line gas ($\sim$1000~\cmt\ compared to $\sim$500--900~\cmt). The line ratio enhancements described above do not show up in the summed spectrum since they have been diluted by the surrounding spaxels.

\begin{figure}
\centering
\includegraphics[width=\textwidth]{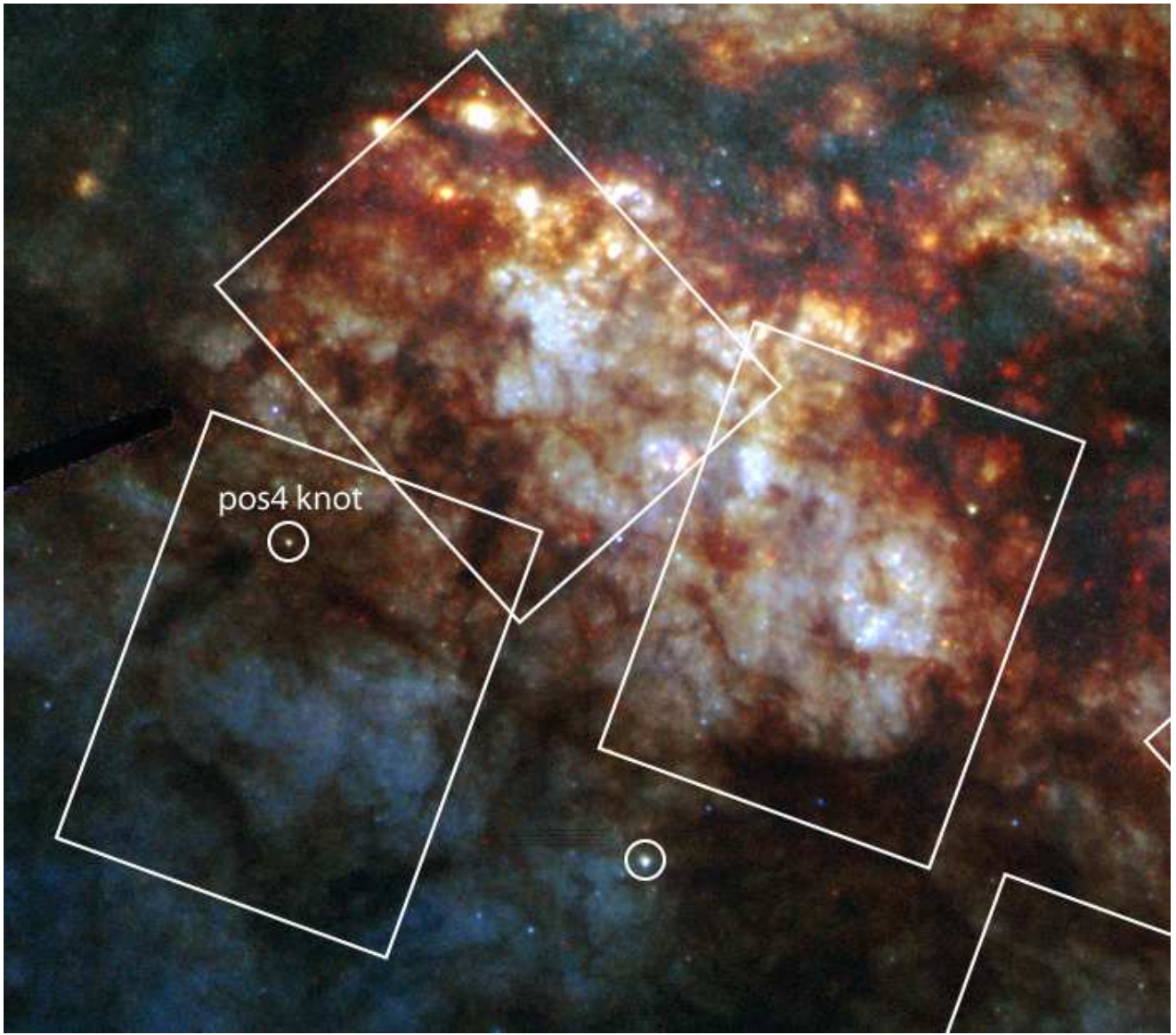} 
\caption{\textit{HST} ACS/HRC colour composite (F330W, F435W, F550M, F814W) image with the position 4 knot and the other similar objects identified, together with the GMOS IFU FoVs. \textit{(A color version of this figure is available in the online journal.)} }
\label{fig:knot_finder}
\end{figure}

Inspection of archived \textit{HST}/ACS HRC images (Prop ID: 10609, PI: Vacca) reveals one other similar nearby point-like source (hereafter source 2), the location of which, together with the position 4 knot, is identified on a F330W ($\sim$$U$), F435W ($\sim$$B$), F550M ($\sim$$V$) and F814W ($\sim$$I$) HRC colour composite in Fig.~\ref{fig:knot_finder}. Measurement of their sizes \citep[using {\sc ishape};][]{larsen04} show that they are both unresolved (FWHM $\lesssim$ $0\farcs06$) in each of the ACS/HRC images, including F658N ($\sim$H$\alpha$). For both objects, we find no evidence for variability ($>$0.2~mag) in any of the aforementioned filters over the epoch range September 1995 to March 2006 (the range covered by all the \textit{HST} WFPC2 and ACS images available in the archive). Absolute F555W magnitudes ($M_{V}$) and colours are given in Table~\ref{tbl:knot_photom}. No evidence of the object is found in any of the available \textit{HST}/NICMOS NIC1/2 images (F160W, F164N, F187N, F190N).

Considering our spectroscopic and photometric evidence and their unresolved nature in the \textit{HST} imaging, there are a number of plausible explanations for these sources. We can immediately dismiss the position 4 knot as being a background object based on its radial velocity, it only having a difference of $\sim$140~\kms\ relative to the systemic velocity of M82. Furthermore, since it does not exhibit an enhanced density compared to its surroundings (Fig.~\ref{fig:elecdens_C2}), does not show up in any of our IFU line or continuum flux maps, and does not correlate with any known radio source at this location \citep{huang94, wills97, mcdonald02}, it is unlikely to be a supernova remnant.

We find the most likely explanation is that the position 4 knot is a hot stellar object immersed in an unresolved ionized circumstellar nebula. The FWHM of the broad component then represents the expansion of the nebula of $\sim$165~\kms. We derive an absolute magnitude of $M_{V}\sim -5.2$ which increases to $\sim$$-7$ if we assume a $V$-band extinction of $\sim$2~mags, suggesting a massive star. The photometry shows that the star is fairly red but this may be due to a dusty circumstellar nebula. The high [N\two]/H$\alpha$ ratio suggests that the nebula may be enriched in nitrogen. All these findings lead us to speculate that the position 4 knot may be a \textit{luminous blue variable} (LBV) star \citep{humphreys94}. These objects are usually surrounded by nitrogen-rich circumstellar material which has been ejected as a result of an unstable event in the past. These stars can spend many years in a hot, quiescent phase and have similar absolute magnitudes to the position 4 knot. The large radial velocity of the emission from this object is intriguing; it might indicate that the star was ejected from the cluster in which it was formed, thus explaining how such a star is found so far from the main star forming complexes. Measuring a projected distance of the star from the nearest star-forming region is $\sim$5--6$''$ ($\sim$90--100~pc), and even assuming a conservative estimate of the age of the LBV (a 60 $M_{\odot}$ mass LBV would be $\lesssim$4~Myr old), the predicted velocity is quite plausible ($<$50\kms).

Source 2, for which we only have photometry measurements, is brighter and redder than the position 4 knot (Table~\ref{tbl:knot_photom}). Although it is also unresolved on the \textit{HST}/ACS HRC images and also does not correlate with any known radio sources. Spectroscopic measurements, well within the capabilities of 8--10~m class telescopes, are needed to ascertain the true nature of these objects.

\section{Summary}\label{sect:summary}

By combining both detailed ($\sim$$0\farcs7$) observations of M82's starburst core and inner wind made with the Gemini GMOS-IFU, and wider-field observations of the inner $\sim$2~kpc of the disk made with the WIYN DensePak instrument, we have been able to build up a comprehensive picture of the state of the ionized gas within this galaxy. To do this we split the study, first focussing on the ionized gaseous and stellar dynamics of the M82 starburst core and inner disk \citepalias{westm09a}, then on the derived nebular properties of the ionized gas (this paper). 

The dynamics showed that the ionized gas within the starburst is dynamically highly complex with much localised line splitting, and many overlapping expanding structures located at different radii. An offset of $\sim$12$^{\circ}$ between the rotation axis of the stars and gas provides further evidence of the effects of the interaction with M81 that is thought to have initially triggered the starburst. We have identified an underlying broad component in the optical nebular emission lines throughout the inner disk, and following the recent work by \citet{westm07a, westm07b} and \citetalias{westm07c}, have associated it with turbulent layers on the surfaces of interstellar clouds set up by the impact of high energy photons and fast-flowing winds from the embedded young star clusters, and hence the sites of mass loading within the wind outflow.

Maps of the [S\two]-derived electron densities show that in both the narrow and broad component gas, significant small-scale local variations are present, and that the peaks are of a few 1000~\cmt\ implying very large interstellar pressures (P/k\,$>$\,$10^7$). Despite these variations and a dramatic fall-off in thermal pressure with height, we still find an impressive uniformity in the line ratios and excitations within the whole inner disk region, meaning that the excitation of the gas within the starburst is extremely uniform. That the ratios are consistent with pure photoionization suggests that shocks must not play an important role in setting the gas ionization within the disk. Our results thus join with those of previous 
studies in showing that the inner region of M82 has many of the characteristics of a single 
supergiant Str\"{o}mgren sphere.

These findings, when put together with the body of literature built up over the last decade on the state of the ISM in M82, imply that the ISM is highly fragmented into a range of clouds from small/dense clumps with low filling factors ($<$1~pc, $n_{\rm e}$\,$\gtrsim$\,$10^4$~\cmt) to larger filling factor, less dense gas. The most compact clouds seem to be found in the clump cores, whereas the cloud sizes in the inter-complex region are larger. These dense clouds are bathed with an intense radiation field and embedded in an extensive high temperature ($T$\,$\gtrsim$\,$10^6$~K) ISM that is a product of the high star formation rates in the starburst zones of M82. The near-constant ionization state of the $\sim$10$^4$~K gas throughout the M82 starburst zone can be explained as a consequence of the small cloud sizes, which allow the gas conditions to respond quickly to any changes.

In our model, schematically represented in the left-centre panel of Fig.~\ref{fig:super_fig}, the broad component (C2) is only emitted from turbulent mixing layers on cloud surfaces in the direct line-of-sight of a wind source. The large number of compact clouds mixed in with the wind sources of course provides an ideal environment for copious emission of these broad lines. In contrast, the narrow components are emitted by relatively quiescent 10$^4$~K gas within the M82 Str\"{o}mgren sphere, including photoionization fronts beneath the TMLs on the surfaces of dense clouds, ionized gas clouds which are not in the direct line-of-sight of an ionizing/wind source (i.e.\ wind shadowed), and general diffuse ionized gas.

Previous observations of the bright region A super star cluster M82-A1 \citep{smith06} indicate that its wind has been stalled by the high interstellar pressures surrounding it. If this cluster is surrounded by a static spherical shell then it would be difficult to undertsand how it, and possibly other massive star clusters, contribute to launching the M82 superwind. However, we can now see that the rapid fall in densities from the north of our IFU fields into the cores of complexes A and C and beyond shows that the conditions towards the south of the complexes (their outward surfaces) may be much more favourable to the escape of the individual cluster winds. Clusters towards the outside surfaces of the starburst complexes may in fact be the primary drivers of the superwind, not clusters such as M82-A1 \citep{smith06} which our maps show is embedded in one of the densest parts of the starburst. This idea is consistent with the asymmetric distribution of extraplanar material in M82: a preponderance of blue sources are found to the south of of the disk by \citet{davidge08}, and a clearly dominant southern blowout region is seen in the UV \citep{hoopes05}.

By extracting major axis position-velocity plots, we showed in \citetalias{westm09a} that the broad-line emitting gas rotates, on average, at a slower velocity than the narrow-line gas. The identification of a discrete channel within the outflow allowed us to explain this velocity difference: if the broad component, representing turbulent gas stirred up by the interaction between the hot, fast wind and cooler gas, becomes more dominant as the gas becomes increasingly perturbed and entrained along the length of the channel, then its emission will be weighted towards regions further out along the flow compared to the narrow component, whose emission is weighted towards the base of the outflow channel.

This idea is supported by the findings of this paper, where we saw that the C1 densities peak not in the starburst complexes themselves, but further in within the major axis dust lane, whereas C2 is at its densest within complexes A and C (i.e.\ further out) and exhibits a slower fall off (with height) than C1. This result, when combined with the fact that the C2/C1 flux ratio increases, and the C2 line widths, although already high in the nuclear regions, also increase (peaking at the location of the minor axis velocity inflection points) imply that, within the energy injection zone (as defined by the minor axis velocity inflection points), the turbulent motions as traced by the broad C2 component appear to play an increasing role with height. This behavior parallels the situation that we found in NGC 1569, and is likely to result from the operation of similar physical processes in both galaxies \citep{westm08}. If the broad component indeed originates from the wind-clump interaction layers where material is also ablatively stripped from the clouds, then although the majority of the mass loading might take place within the starburst complexes, the region of mass loading must extend far out into the inner wind.
  
The narrow-line (C1 and C3) H$\alpha$ position-velocity plots extracted from GMOS IFU position 5 have allowed us to look in more detail at the aforementioned outflow channel. The channel appears as a coherent, expanding cylindrical structure extending along the whole length of position 5, increasing slightly in physical width from the north to south from $\sim$2--3$''$ (35--50~pc), and maintaining an approximately constant (subsonic) expansion velocity of $\sim$60~\kms. Velocities of the broad C2 line suggest that it originates within the space between the two walls. The walls of the channel are defined by peaks in the C2 densities (700--1000~\cmt) and troughs in the C1 densities ($\sim$200~\cmt), whereas the centre of the channel is defined by higher C1 densities ($\gtrsim$500~\cmt) that decrease with radius. In our model, the channel cavity contains the hot wind fluid that, as it flows outward, interacts with the cooler, denser walls of the channel and with clumps of entrained material to produce broad-line emission, while the walls themselves emit primarily the narrow lines C1 and C3. Hints are seen that there may be multiple similar channels within the outflow.

Finally, we have argued that a point-like knot identified in GMOS position 4, exhibiting blueshifted ($\sim$140~\kms), broad ($\lesssim$350~\kms) H$\alpha$ emission and enhanced [S\two]/H$\alpha$ and [N\two]/H$\alpha$ ratios, is most likely a \textit{luminous blue variable} (LBV) type object, that may have been ejected from the cluster in which it was formed.


\section*{Acknowledgments}
We would like to thank the WIYN Observatory staff for excellent support of this programme. JSG's research was partially funded by the National Science Foundation through grant AST-0708967 to the University of Wisconsin-Madison.

Based on observations obtained at the Gemini Observatory, which is operated by the Association of Universities for Research in Astronomy, Inc., under a cooperative agreement with the NSF on behalf of the Gemini partnership: the National Science Foundation (United States), the Science and Technology Facilities Council (United Kingdom), the National Research Council (Canada), CONICYT (Chile), the Australian Research Council (Australia), Minist\'{e}rio da Ci\^{e}ncia e Tecnologia (Brazil) and SECYT (Argentina).

\bibliographystyle{apj}
\bibliography{/Users/msw/Documents/work/references}





\clearpage

\end{document}